\documentclass[10pt,twocolumn,pra,aps,showkeys,showpacs]{revtex4-1}

\usepackage[T1]{fontenc}
\usepackage[latin9]{inputenc}
\usepackage{mathrsfs}
\usepackage{amstext}
\usepackage{amsmath}
\usepackage{amssymb}
\usepackage{graphicx}
\usepackage[unicode=true,
 bookmarks=false,
 breaklinks=false,pdfborder={0 0 1},colorlinks=true]
 {hyperref}
 
\usepackage[svgnames]{xcolor}
\usepackage{pdfcolmk}
\usepackage{bm}

\begin{document}
\title{Non-Abelian geometric phases in periodically driven systems}



\author{Viktor Novi\v{c}enko}
\email[]{viktor.novicenko@tfai.vu.lt}
\homepage[]{http://www.itpa.lt/~novicenko/}
\affiliation{Institute of Theoretical Physics and Astronomy, Vilnius University,
Saul\.{e}tekio Ave.~3, LT-10257 Vilnius, Lithuania}

\author{Gediminas Juzeli\={u}nas}
\email[]{gediminas.juzeliunas@tfai.vu.lt}
\homepage[]{http://www.itpa.lt/~gj/}
\affiliation{Institute of Theoretical Physics and Astronomy, Vilnius University,
Saul\.{e}tekio Ave.~3, LT-10257 Vilnius, Lithuania}


\date{\today}

\begin{abstract}

We consider a periodically driven quantum system described by a Hamiltonian which is the
product of a slowly varying Hermitian operator $V\left(\boldsymbol{\lambda}\left(t\right)\right)$ and a dimensionless periodic function with zero average.
We demonstrate that the adiabatic evolution of the system within a fully
degenerate Floquet band is accompanied by non-Abelian (non-commuting) geometric phases
appearing when the slowly varying parameter $\boldsymbol{\lambda}=\boldsymbol{\lambda}\left(t\right)$
completes a closed loop. The geometric phases can have significant
values even after completing a single cycle of the slow variable. Furthermore, there are no dynamical
phases masking the non-Abelian Floquet geometric phases, as the former average to zero over an
oscillation period. This can be used to precisely control the evolution of quantum systems, in
particular for performing qubit operations. The general formalism is illustrated by analyzing a spin
in an oscillating magnetic field with arbitrary strength and a slowly changing direction.

\end{abstract}


\maketitle


\section{Introduction\label{sec:Introduction}}

Topological and many-body properties of physical systems can be enriched
by applying a periodic driving~\cite{Oka2009,Kitagawa2010,Kitagawa2011,Galitski2011NP,kolovsky11,creffield13comment,Rudner2013,Nathan15NJP,Weinberg17PR,Kolodrubetz18PRL,Sacha18RMP}.
This extends to a wide range of condensed matter~\cite{Oka2009,Galitski2011NP,Kitagawa2011,Galitski13PRB,tong13majorana,grushin14,usaj14,quelle15,Klinovaja2017,Gavensky18PRB},
photonic~\cite{Haldane:2008cc,Rechtsman:2013fe,Mukherjee17Ncommun,Jorg17NJP,Mukherjee18NCommun}
and ultracold atom~\cite{zenesini09,Dalibard2011,Aidelsburger:2011,struck11,Arimondo2012,struck12,Hauke:2012,Windpassinger2013RPP,Aidelsburger:2013,hauke13,Struck:2013,Ketterle:2013,Anderson2013,Xu2013,Galitski2013,Goldman2014RPP,Aidelsburger14NP,atala14,Jotzu2014,Zhai2014-review,Kennedy15,Flaschner16,Budich16,Nagerl2016,jimenez15,nascimbene15,perez-piskunow15,Luo16Sci_Rep,Eckardt17RMP,Schneider17PRL,Chin18PRL,Weitenberg18arXiv,Shteynas18arXiv}
systems. For example, the periodic driving can induce a non-staggered
synthetic magnetic flux~\cite{kolovsky11,Aidelsburger:2013,Ketterle:2013,Aidelsburger14NP,Goldman2014RPP,Kennedy15,Eckardt17RMP}
or facilitate the realization of the Haldane model \cite{Oka2009,Jotzu2014}
for ultracold atoms in optical lattices. To deal with periodically
driven  quantum systems, it is convenient to describe their long-term dynamics
in terms of an effective time-independent Floquet Hamiltonian. In
that case fast oscillations of the system within a driving period
are represented by a micromotion operator. If the driving frequency
exceeds other characteristic frequencies of the system, the Floquet
Hamiltonian and the micromotion operator can be expanded in inverse
powers of the driving frequency~\cite{Rahav03,Goldman2014,goldman15resonant,Eckardt2015,Bukov2015,itin15,Mikami16PRB,Heinisch16,holthaus16tutorial,Eckardt17RMP}. 

It is quite common that the periodic driving changes in time. For
example, in typical ultracold atom experiments one ramps up the periodic
driving from zero to a stationary regime \cite{Esslinger17PRA}. In
the previous paper \cite{Novicenko2017} we have considered such a
situation where a quantum system is subjected to a periodic driving
which changes slowly in time. High frequency expansions have been
obtained for the effective Hamiltonian and for the micromotion operators showing
that these operators change in time because of the changes in the
periodic driving \cite{Novicenko2017}. Furthermore the expanded effective
Hamiltonian contains an extra second order term  appearing due to the
slow changes in the periodic driving  \cite{Novicenko2017}.  
This can provide non-Abelian (non-commuting) geometric phases for periodically
driven systems.

The high frequency expansion was applied to a spin in a fast oscillating magnetic field with a slowly changing amplitude \cite{Novicenko2017}. If the magnetic field slowly changes its direction performing a cyclic evolution in 3D space, a non-Abelian geometric phase appears after the slow variable (the magnetic field amplitude) completes a cycle. Yet the acquired phase represents a small second order correction, so the slow variable should complete many cycles to accumulate a substantial geometric phase. This is because the high frequency expansion is applicable only if the driving strength is small compared to the driving frequency.   The current analysis does not rely on such an approximation for the periodic driving. We show that the system can acquire substantial geometric phases even if the slow variable completes just a single cycle.    

 We study a periodically driven quantum system characterized
by a Hamiltonian which is a product of a slowly varying Hermitian
operator and a fast oscillating periodic function with a zero average.
 We transform the equations of motion to a new representation by applying a time dependent unitary transformation. 
The transformation eliminates the original Hamiltonian in the equations of motion, and there is an extra term due the slow changes in the periodic driving.  
Neglecting the latter term, individual Floquet bands are completely degenerate. The slow changes of the driving couples the Floquet states. 
 We apply the adiabatic approximation by neglecting the coupling between different Floquet bands separated by the driving frequency times an integer.
This is equivalent to the zero-order of the high-frequency expansion \cite{Novicenko2017} of the Floquet effective Hamiltonian in the transformed representation. 
It is demonstrated that the adiabatic evolution of the system within an individual  degenerate Floquet band is accompanied by non-Abelian
geometric phases which can be sufficiently 
large even after completing a single cycle of the slow variable. 
Furthermore, there are no dynamical
phases masking the non-Abelian Floquet geometric phases, as the former
average to zero over an oscillation period. This can be used for precisely
controlling the evolution of quantum systems, in particular for performing
qubit operations.

The paper is organized as follows. In the subsequent Sec.~\ref{sec:Formulation}
we define a periodically driven system with a slowly modulated driving
and go to a new representation via a time-dependent unitary transformation. 
In Sec.~\ref{sec:Adiabatic-approach} we
consider the adiabatic evolution
of the system within  an individual Floquet band, and show that the evolution is accompanied by non-Abelian geometric phases.
In Sec.~\ref{sec:Analysis-of operator W}
we analyze the operator responsible for the geometric
phases and provide explicit expressions for this operator in specific situations.
Section~\ref{sec:Spin-in-oscillating} illustrates the general formalism
by analyzing a spin in an oscillating magnetic field with arbitrary
strength and a slowly changing direction. The concluding Sec.~\ref{sec:Concluding-remarks}
summarizes the findings. Technical details of some calculations are presented in the two Appendices \ref{sec:Appendix A} and \ref{sec:Appendix B}.

\section{Periodically driven system with a modulated driving\label{sec:Formulation}}

\subsection{Hamiltonian and equations of motion}

We consider a periodically driven (Floquet) quantum system with a slowly
modulated driving. The system is described by a Hamiltonian which is
the product of a slowly varying Hermitian operator $V\left(\boldsymbol{\lambda}\right)=V\left(\boldsymbol{\lambda}\left(t\right)\right)$
and a fast oscillating dimensionless real function $f\left(\omega t+\theta\right)$
\begin{equation}
H\left(\omega t+\theta,t\right)=V\left(\boldsymbol{\lambda}\left(t\right)\right)f\left(\omega t+\theta\right)\,,\label{eq:H_full}
\end{equation}
where $\omega$ is the oscillation frequency and $\theta$ defines the phase of the oscillations. 
The operator $V\left(\boldsymbol{\lambda}\right)$
depends on time via a set of slowly varying parameters $\boldsymbol{\lambda}=\boldsymbol{\lambda}\left(t\right)=\left\{ \lambda_{\mu}\left(t\right)\right\} $
which change little over the driving period $T=2\pi/\omega$, the subscript $\mu$ specifies individual slowly varying parameters. Here also $f\left(\omega t+\theta\right)$ is taken to be a $2\pi$ periodic
function $f\left(\omega t+\theta+2\pi\right)=f\left(\omega t+\theta\right)$
with an amplitude of the order of unity and zero average: $\intop_{0}^{2\pi}f\left(\theta^{\prime}\right)\mathrm{d}\theta^{\prime}=0$.
Therefore the Fourier expansion of the Hamiltonian
\begin{equation}
H\left(\theta^{\prime},t\right)=\sum_{m=-\infty}^{\infty}H^{\left(m\right)}\left(t\right)e^{\mathrm{i}m\theta^{\prime}} \; \mathrm{with}\quad \theta^{\prime}=\omega t+\theta
\label{eq:H-periodic-expansion}
\end{equation}
does not contain a zero frequency component, ${H^{\left(0\right)}\left(t\right)=0}$, while other components are
\begin{equation}
H^{\left(m\right)}\left(t\right)=V\left(\boldsymbol{\lambda}\left(t\right)\right)f^{\left(m\right)}\,,\label{eq:H^m}
\end{equation}
with
$f^{\left(m\right)}=\intop_{0}^{2\pi}f\left(\theta^{\prime}\right)e^{-\mathrm{i}m\theta^{\prime}}\mathrm{d}\theta^{\prime}$.

An example of such a system is a spin in a magnetic field $\mathbf{B}\left(t\right)f\left(\omega t+\theta\right)$ \cite{Novicenko2017}
with a fast oscillating amplitude ${\propto f\left(\omega t+\theta\right)}$
and a slowly changing direction $\propto\mathbf{B}\left(t\right)$, where $\mathbf{B}\left(t\right)$
plays the role of $\boldsymbol{\lambda}\left(t\right)$. In that case the slowly varying part of the
Hamiltonian is given by
\begin{equation}
V\left(\mathbf{B}\left(t\right)\right)=g_{F}\mathbf{F}\cdot\mathbf{B}\left(t\right)\,,\label{eq:V-spin}
\end{equation}
where $g_{F}$ is a gyromagnetic factor, $\mathbf{F}=F_{1}\mathbf{e}_{x}+F_{2}\mathbf{e}_{y}+F_{3}\mathbf{e}_{z}$
is a spin operator with Cartesian components satisfying the usual
commutation relations $\left[F_{s},F_{q}\right]=\mathrm{i}\hbar\epsilon_{sqr}F_{r}$.
Here $\epsilon_{sqr}$ is the Levi-Civita symbol, and a summation over
a repeated Cartesian index $r=x,y,z$ is implied. 

A state-vector $\left|\phi\left(t\right)\right\rangle $ of the system
belongs to the Hilbert space $\mathscr{H}$ and obeys the time-dependent
Schr\"{o}dinger equation (TDSE):
\begin{equation}
\mathrm{i}\hbar\frac{\partial}{\partial t}\left|\phi\left(t\right)\right\rangle =V\left(\boldsymbol{\lambda}\left(t\right)\right) 
f\left(\omega t+\theta\right)\left|\phi\left(t\right)\right\rangle \,.\label{eq:Schroed-time-dep-H}
\end{equation}
Previously a general perturbative analysis was carried out to deal
with the evolution of a periodically driven system with a modulated
driving using the Floquet extended-space approach \cite{Novicenko2017}.
Such a perturbative treatment is generally valid if matrix elements
of the Fourier components of the periodic Hamiltonian are
small compared to the driving frequency:
\begin{equation}
\left|H_{\alpha\beta}^{\left(m\right)}\right|\ll\hbar\omega\,, \quad \mathrm{and\,\, hence} \quad \left|V_{\alpha\beta}\right|f^{\left(m\right)}\ll \hbar\omega\,,
\label{eq:High Frequency condition general}
\end{equation}
where the subscripts $\alpha$ and $\beta$ are used to label the matrix element of the operators. Furthermore,
$H_{\alpha\beta}^{\left(m\right)}$ should  change sufficiently slowly over the driving period.

\subsection{Transformed representation } 

\begin{figure}[h]
\includegraphics[width=0.45\textwidth]{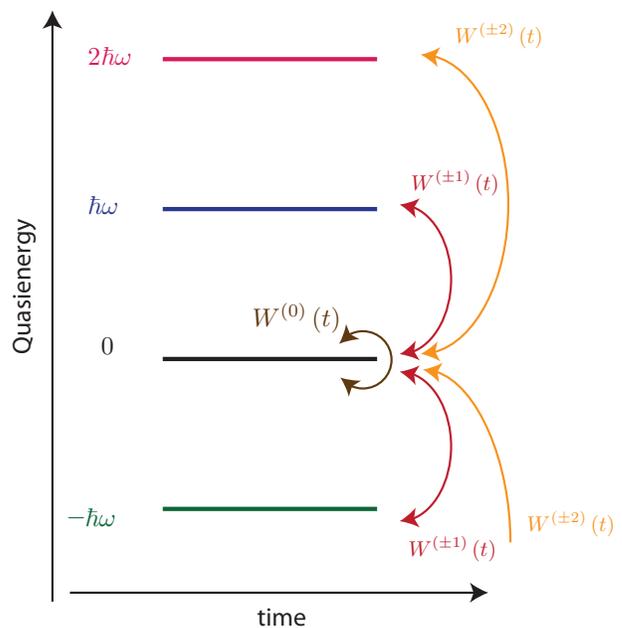} 
\caption{\label{fig:Floquet_level_loop}(Color online) Schematic representation of coupling between the Floquet bands in the transformed representation described by Eqs.~(\ref{eq:psi^n-equation})-(\ref{eq:K-definition}).  The
operators $W^{\left(m\right)}\left(t\right)$
with $m\protect\ne0$ describe coupling between different  Floquet
manifolds, whereas  the operator $W^{\left(0\right)}\left(t\right)$ couples the states belonging to the same degenerate Floquet band. The latter $W^{\left(0\right)}\left(t\right)$ provides non-Abelian the geometric phases  for the adiabatic motion of the system within the same Floquet manifold. }
\end{figure}

In what follows we will consider the dynamics of the system when the
weak driving condition~(\ref{eq:High Frequency condition general})
does not necessarily hold. For this, we will go to another representation via a unitary operator
which eliminates the original Hamiltonian in the transformed equations of motion. Such a unitary operator reads 
\begin{equation}
R\left(\omega t + \theta,\boldsymbol{\lambda}\left(t\right) \right)=\exp\left[-\mathrm{i}\frac{\mathcal{F}\left(\omega t + \theta\right)}{\hbar\omega}V\left(\boldsymbol{\lambda}\left(t\right)\right)\right]\,,
\label{eq:R-Definition}
\end{equation}
where $\mathcal{F}\left(\theta^{\prime}\right)$ is a primitive function
of $f\left(\theta^{\prime}\right)$ with zero average: 
\begin{equation}
\mathrm{d}\mathcal{F}\left(\theta^{\prime}\right)/\mathrm{d}\theta^{\prime}=f\left(\theta^{\prime}\right)\quad\mathrm{and}\quad\intop_{0}^{2\pi}\mathcal{F}\left(\theta^{\prime}\right)\mathrm{d}\theta^{\prime}=0\,.\label{eq:F derivative}
\end{equation}
The calligraphy letter $\mathcal{F}$ is used to avoid confusion
with the spin operator $\mathbf{F}$ featured in Eq.~(\ref{eq:V-spin}). 

 The transformed state-vector 
\begin{equation}
\left|\psi\left(t\right)\right\rangle =R^{\dagger}\left(\omega t+\theta,t\right)\left|\phi\left(t\right)\right\rangle \,,\label{eq:|chi_theta>}
\end{equation}
obeys the TDSE 
\begin{equation}
\mathrm{i}\hbar\frac{\partial}{\partial t}\left|\psi\left(t\right)\right\rangle =W\left(\omega t+\theta,t\right)\left|\psi\left(t\right)\right\rangle \,,\label{eq:Schroed-time-dep-K_R}
\end{equation}
with 
\begin{equation}
W\left(\theta^{\prime},t\right)=-\mathrm{i\hbar}R^{\dagger}\left(\theta^{\prime},\boldsymbol{\lambda}\left(t\right)\right)\partial R\left(\theta^{\prime},\boldsymbol{\lambda}\left(t\right) \right)/\partial t\,,\label{eq:W-definition}
\end{equation}
where the partial derivative $\partial R/\partial t$ is calculated
for a fixed value of the variable $\theta^{\prime}=\omega t+\theta$. 
Alternatively,
the transformed Hamiltonian can be represented as:
\begin{equation}
W\left(\theta^{\prime},t\right)=\dot {\lambda}_{\mu} A_{\mu}\left(\theta^{\prime},\boldsymbol{\lambda}\right)\,,
\label{eq:W-definition_A}
\end{equation}
where summation over repeated indices $\mu$ is implied, and
\begin{equation}
A_{\mu}\left(\theta^{\prime},\boldsymbol{\lambda}\right)=-i\hbar R^{\dagger}\left(\theta^{\prime},\boldsymbol{\lambda}\right)\frac{\partial R\left(\theta^{\prime},\boldsymbol{\lambda}\right)}{\partial\lambda_{\mu}} 
\label{eq:A_mu}
\end{equation}
is the $\mu$th component of the vector potential $\mathbf{A}\left(\theta^{\prime},\boldsymbol{\lambda}\right)$.

In this way, the transformation $R\left(\theta^{\prime},\boldsymbol{\lambda}\left(t\right)\right)$ eliminates the original Hamiltonian
(\ref{eq:H_full}) in the transformed
equation of motion~(\ref{eq:Schroed-time-dep-K_R}). The new Hamiltonian
$W \left(\omega t+\theta,t\right)$ given by Eqs.~(\ref{eq:W-definition})-(\ref{eq:W-definition_A}) is due to the slow temporal changes
of the variable $\boldsymbol{\lambda} = \boldsymbol{\lambda}\left(t\right)$ entering the transformation~(\ref{eq:R-Definition}).  Therefore
the transformed Hamiltonian $W\left(\omega t+\theta,t\right)$ can be arbitrarily small 
even if the original weak driving (high frequency) condition 
(\ref{eq:High Frequency condition general}) is violated. 
In particular, one has
$W\left(\omega t+\theta,t\right)=0$ for a pure periodic driving where $\boldsymbol{\lambda}\left(t\right)$ is constant. 

The evolution of the transformed state-vector can be represented as
\begin{equation}
\left|\psi\left(t\right)\right\rangle =U\left(t,t_{0}\right)\left|\psi\left(t_{0}\right)\right\rangle,
\label{eq:psi}
\end{equation}
with
\begin{equation}
U\left(t,t_{0}\right)={\cal T}\exp\left[-\frac{\mathrm{i}}{\hbar}\int_{t_{0}}^{t}W\left(\omega t^{\prime}+\theta,t^{\prime}\right)\mathrm{d}t^{\prime}\right],
\label{eq:psi-t-solution}
\end{equation}
where ${\cal T}$ indicates the time ordering and $t_{0}$ is an initial
time. The operator $W$ determining the evolution of the transformed
state vector will be analyzed in Sec.~\ref{sec:Analysis-of operator W}.

Like the original Hamiltonian $H\left(\omega t+\theta,t\right)$,
the transformed Hamiltonian $W\left(\omega t+\theta,t\right)$ is
$2\pi$ periodic with respect to the first variable and thus can be
expanded in a Fourier series with respect to the fast variable:
\begin{equation}
W\left(\omega t+\theta,t\right)=\sum_{n=-\infty}^{\infty}W^{\left(n\right)}\left(t\right)e^{\mathrm{i}n\left(\omega t+\theta\right)}\,,
\label{eq:W-periodic-expansion}
\end{equation}
where
\begin{equation}
W^{\left(n\right)}\left(t\right)=\frac{1}{2\pi}\intop_{0}^{2\pi}W\left(\theta^{\prime},t\right)e^{-\mathrm{i}n\theta^{\prime}}\mathrm{d}\theta^{\prime}\,.
\label{eq:W-n}
\end{equation}

Expanding also the transformed state-vector $\left|\psi\left(t\right)\right\rangle \equiv\left|\psi\left(\omega t+\theta,t\right)\right\rangle$:
\begin{equation}
\left|\psi\left(\omega t+\theta,t\right)\right\rangle =\sum_{n=-\infty}^{\infty}\left|\psi^{\left(n\right)}  \left(t\right)\right\rangle e^{\mathrm{i}n\left(\omega t+\theta\right)}\,,
\label{eq:psi-expansion}
\end{equation}
the TDSE (\ref{eq:Schroed-time-dep-K_R}) provides the following equation for the slowly changing Fourier components $\left|\psi^{\left(n\right)}  \left(t\right)\right\rangle$:
\begin{equation}
\mathrm{i} \hbar \frac{\partial} {\partial t} \left|\psi^{\left(n\right)}  \left(t\right)\right\rangle  =  
\sum_{m=-\infty}^{\infty} K_{nm}  \left|\psi^{\left(m\right)}  \left(t\right)\right\rangle
\,,
\label{eq:psi^n-equation}
\end{equation}
where
\begin{equation}
K_{nm} = n\hbar  \omega \delta_{nm} + W^{\left(n-m\right)}\left(t\right)
\,
\label{eq:K-definition}
\end{equation}
are the matrix elements of the extended space Floquet Hamiltonian $\hat{K}\left(t\right)$  slowly changing in time  \cite{Novicenko2017}. Its off diagonal terms $K_{nm}=W^{\left(n-m\right)}\left(t\right)$ with $n\ne m$ describe the coupling between different Fourier components (different Floquet bands)  $\left|\psi^{\left(n\right)}  \left(t\right)\right\rangle$ and  $\left|\psi^{\left(m\right)}  \left(t\right)\right\rangle$  due to the changes of the periodic driving. The diagonal elements $K_{nn} = n\hbar \omega+W^{\left(0\right)}\left(t\right)$ contain the energy of the $n$th Floquet manifold $n\hbar \omega$ and an extra operator $W^{\left(0\right)}\left(t\right)$ emerging due to the changes of the periodic driving. Figure~\ref{fig:Floquet_level_loop} illustrates the coupling between  different Floquet manifolds ($n\ne m$) and within 
of the same Floquet bands ($n=m$). Neglecting all coupling terms $W^{\left(n-m\right)}\left(t\right)$ in Eq.~(\ref{eq:K-definition}), the eigenstates of the operator $\hat{K}\left(t\right)$ are completely degenerate within individual Floquet bands with quasi-energies $n\hbar \omega$ shown by horizontal lines in Fig.~\ref{fig:Floquet_level_loop}. When the effects due to the changes of the periodic driving are included, the emerging operator $W^{\left(0\right)}\left(t\right)$  provides the Floquet geometric phases for the adiabatic motion within a single  degenerate Floquet manifold.   We will consider this issue in more details in the next Section.

\section{Adiabatic approach\label{sec:Adiabatic-approach} }

\subsection{Effective evolution operator in transformed representation}

We are interested in a situation where $V\left(\boldsymbol{\lambda}\left(t\right)\right)$ changes
sufficiently slowly, so the matrix elements of the Fourier components
of $W\left(\omega t + \theta,t\right)$ are smaller than the driving frequency
\begin{equation}
\left|W_{\alpha\beta}^{\left(n\right)}\right|\ll\hbar\omega\,,\label{eq:High Frequency condition general_for_W}
\end{equation}
and also change sufficiently smoothly. 
The condition (\ref{eq:High Frequency condition general_for_W}) has the same form as the original condition (\ref{eq:High Frequency condition general}) with $H$ replaced by $W $. Since the transformed Hamiltonian
$W \left(\omega t+\theta,t\right)$ given by Eqs.~(\ref{eq:W-definition})-(\ref{eq:W-definition_A}) is due to the temporal changes
of $V \left(\boldsymbol{\lambda}\left(t\right)\right)$, the condition (\ref{eq:High Frequency condition general_for_W})
relies on the slow changes of the periodic driving rather that on its
weakness. Therefore Eq.~(\ref{eq:High Frequency condition general_for_W})
can hold even if the matrix elements $\left|H_{\alpha\beta}^{\left(n\right)}\right|$
exceed $\hbar\omega$ and thus there is a violation of the original high-frequency
requirement~(\ref{eq:High Frequency condition general}).

Applying the condition (\ref{eq:High Frequency condition general_for_W}), the evolution of the transformed state-vector can be described by means of the slowly changing Floquet effective Hamiltonian $W_{\textrm{eff}}\left(t\right) $ 
 expanded in the inverse powers of the driving frequency $\omega^{(-n)}$ (with $n\ge0$)  
 \cite{Novicenko2017}. 
 The adiabatic approximation is obtained by keeping only the zero order term of the effective Hamiltonian $W_{\textrm{eff}(0)}\left(t\right)=W^{\left(0\right)}\left(t\right)$ in the transformed TDSE~(\ref{eq:Schroed-time-dep-K_R}). 
In other words, due to condition~(\ref{eq:High Frequency condition general_for_W}) one neglects off diagonal terms $W^{\left(n-m\right)}\left(t\right)$ with $n\ne m$ which describe the coupling between different Floquet bands in Eq.~(\ref{eq:psi^n-equation}), 
as illustrated in Fig.~\ref{fig:Floquet_level_loop}.
This is equivalent to the time averaging of the transformed Hamiltonian $W\left(\omega t + \theta,t\right)$ over fast oscillations. Consequently,  
the evolution operator~(\ref{eq:psi-t-solution}) can be replaced by the effective operator for the adiabatic dynamics, $U\left(t,t_{0}\right)\approx U_{\textrm{eff}(0)}\left(t,t_{0}\right)$, with
\begin{equation}
U_{\textrm{eff}(0)}\left(t,t_{0}\right)={\cal T}\exp\left[-\frac{\mathrm{i}}{\hbar}\int_{t_{0}}^{t}W^{\left(0\right)}\left(t^{\prime}\right)\mathrm{d}t^{\prime}\right]\,,\label{U_eff-definition}
\end{equation}
where the time-ordering ${\cal T}$ is needed 
if the effective Hamiltonian $W^{\left(0\right)}\left(t\right)$ does
not commute with itself at different times, $\left[W^{\left(0\right)}\left(t^{\prime}\right),W^{\left(0\right)}\left(t^{\prime\prime}\right)\right]\ne0$.

\subsection{Non-Abelian geometric phases\label{subsec:Non-Abelian-geometric-phases}}

Calling on
Eq.~(\ref{eq:W-definition_A}) for  $W\left(\omega t + \theta,t\right)$, the adiabatic evolution operator~(\ref{U_eff-definition}) 
can be represented as
\begin{align}
U_{\textrm{eff}(0)}\left(t,t_{0}\right) &= {\cal T}\exp\left[-\frac{\mathrm{i}}{\hbar}\int_{t_{0}}^{t}A_{\mu}^{(0)}\left(\boldsymbol{\lambda}\left(t^{\prime}\right)\right)\mathrm{d}\lambda_{\mu}\left(t^{\prime}\right)\right] \nonumber \\ 
&=\exp\left[\mathrm{i}\Gamma\left(t,t_{0}\right)\right]\,,
\label{U_eff-in-terms-of-vector potential}
\end{align}
where 
\begin{equation}
A_{\mu}^{(0)}\left(\boldsymbol{\lambda}\right)=\frac{1}{2\pi}\intop_{0}^{2\pi}A_{\mu}\left(\theta^{\prime},\boldsymbol{\lambda} \right)\mathrm{d}\theta^{\prime}\,,\label{eq:A^(0)}
\end{equation}
is the zero-frequency Fourier component of $A_{\mu}\left(\omega t + \theta,\boldsymbol{\lambda} \right)$.

The evolution operator $U_{\textrm{eff}(0)}\left(t,t_{0}\right)$ does
not depend on the speed of the change of the parameters $\boldsymbol{\lambda}=\boldsymbol{\lambda}\left(t\right)$. The operator $U_{\textrm{eff}(0)}\left(t,t_{0}\right)$ is defined exclusively by the trajectory along which
the parameters $\boldsymbol{\lambda}=\left\{ \lambda_{\mu}\right\} $
evolve. In particular, if the slowly varying parameters $\boldsymbol{\lambda}=\boldsymbol{\lambda}\left(t\right)$
undergo a cyclic evolution and return to their original values:
$\boldsymbol{\lambda}\left(t\right)=\boldsymbol{\lambda}\left(t_{0}\right)$,
the operator $U_{\textrm{eff}(0)}\left(t,t_{0}\right)=\exp\left(\mathrm{i}\Gamma\right)$
is determined by the geometry of such a closed trajectory. Therefore the operator
$\Gamma$ featured in the exponent describes the geometric
phase acquired during the cyclic evolution. 
When the parameters $\boldsymbol{\lambda}=\left\{ \lambda_{\mu}\right\} $ complete  two consecutive closed loop 
trajectories, the evolution of the system is described by the product of two geometric phase
factors, $\exp\left(i\Gamma_{1}\right)$ and $\exp\left(i\Gamma_{2}\right)$,  corresponding to each closed loop.
If the two factors do not commute,  one arrives at non-Abelian
geometric phases. Thus the present work extends the previous studies of non-Abelian geometric phases  
\cite{Wilczek:1984,Moody1986,Zee1988} to the periodically driven (Floquet) system.
In particular, non-Abelian geometric phases are formed for a spin in 
an oscillating magnetic field with a slowly changing direction, as we will see in Sec.\ref{sec:Spin-in-oscillating}.

 It is noteworthy that the non-Abelian geometric phases emerge because the system is subjected to the periodic driving. The periodic driving
provides degenerate manifolds of Floquet states  (shown by horizontal lines in Fig.~\ref{fig:Floquet_level_loop})
if one neglects $W\left(\omega t+\theta,t\right)$ appearing due the slow changes of the driving. 
Such a degeneracy of the Floquet bands is related to the fact that
the instantaneous eigen-energies of the Hamiltonian~(\ref{eq:H_full}) average to zero
over the driving period. 
The slow change of the driving induces the non-Abelian geometric phase factors represented by $W^{(0)}\left(t\right)=\dot {\lambda}_{\mu} A_{\mu}^{(0)}\left(\boldsymbol{\lambda}\right)$, and which in turn describe the coupling within a degenerate Floquet band in the evolution operator given by Eqs.~(\ref{U_eff-definition}) or (\ref{U_eff-in-terms-of-vector potential}).

We are interested mosly in a situation where the periodically driven quantum system is characterized by a Hilbert space of finite dimension. 
In that case the  number of the Floquet states within each degenerate Floquet band $n$ (represented by horizontal lines in Fig.~\ref{fig:Floquet_level_loop}) 
equals to the number of the basis state-vectors $\left|\alpha\right\rangle $,
i.e. to the dimension of the Hilbert space $\mathscr{H}$. 
In particular, for the spin in the magnetic field described by Eq.~(\ref{eq:V-spin}),
the Hilbert space $\mathscr{H}$ spans all spin projection states
with quantum numbers $m_{F}=-f_{F}\,,-\left(f_{F}-1\right)\,,\ldots\,,+f_{F}$. The number of degenerate states then equals to $2f_{F}+1$ for each Floquet band,
where $f_{F}$ is the spin quantum number.

\subsection{Return to the original representation and Floquet states}

Returning to the original representation 
\begin{equation}
\left|\phi\left(t\right)\right\rangle \equiv\left|\phi\left(\omega t+\theta,t\right)\right\rangle =R\left(\omega t+\theta,\boldsymbol{\lambda}\left(t\right)\right)\left|\psi\left(t\right)\right\rangle \,,\label{eq:phi_inverse transformation}
\end{equation}
the adiabatic evolution of the state vector is given by 
\begin{align}
& \left|\phi\left(\omega t+\theta,t\right)\right\rangle = \nonumber \\
& e^{-\mathrm{i}S\left(\omega t+\theta,t\right)}U_{\textrm{eff}(0)}\left(t,t_{0}\right)e^{\mathrm{i}S\left(\omega t_{0}+\theta,t_{0}\right)}\left|\phi\left(t_{0}\right)\right\rangle \,,\label{eq:phi-t-solution}
\end{align}
where the oscillating Hermitian operator
\begin{equation}
S\left(\omega t+\theta,t\right)=\frac{\mathcal{F}\left(\omega t+\theta\right)}{\hbar\omega}V\left(\boldsymbol{\lambda}\left(t\right)\right)\label{eq:S}
\end{equation}
describes the fast micromotion of the state-vector~(\ref{eq:phi-t-solution}) due to the periodic driving.
The solution $\left|\phi\left(\omega t+\theta,t\right)\right\rangle$ is thus $2\pi$ periodic with
respect to the first variable $\omega t+\theta $. Additionally, $\left|\phi\left(\omega t+\theta,t\right)\right\rangle$
slowly changes with respect to the second variable $t$ due to the
temporal dependence of the operator $V\left(\boldsymbol{\lambda}\left(t\right)\right)$ determining $S\left(\omega t+\theta,t\right)$ and $U_{\textrm{eff}(0)}\left(t,t_{0}\right)$. 

If there is no periodic driving at the initial time, $V\left(\boldsymbol{\lambda}\left(t_{0}\right)\right)=0$,
and the driving is ramped up slowly afterwards, the evolution is not
affected by micromotion due to the ramping of the periodic driving: $S\left(\omega t_{0}+\theta,t_{0}\right)=0$.
If additionally the periodic perturbation is ramped down slowly before
the final time $t$, there is no contribution due to the micromotion
at the final time either: $S\left(\omega t+\theta,t\right)=0$, and
Eq.~(\ref{eq:phi-t-solution}) reduces to
\begin{equation}
\left|\phi\left(\omega t+\theta,t\right)\right\rangle =U_{\textrm{eff}(0)}\left(t,t_{0}\right)\left|\phi\left(t_{0}\right)\right\rangle \,.\label{eq:phi-t-solution-slow ramping}
\end{equation}
In this way, if the driving is ramped up and down slowly, the micromotion
does not contribute to the overall adiabatic evolution of the state vector $\left|\phi\left(\omega t+\theta,t\right)\right\rangle $.
The evolution is determined exclusively by the operator
$U_{\textrm{eff}(0)}\left(t,t_{0}\right)$ containing the non-Abelian
geometric phases. This can be used for precisely controlling the dynamics
of the quantum system, such as for manipulating of qubits. A specific
sequence of ramping up and down of the driving will be discussed in
the Sec. \ref{subsec:Specific-sequence-of} for a spin in the oscillating
magnetic field.

For purely periodic driving where $\boldsymbol{\lambda}\left(t\right)=\boldsymbol{\lambda}$ is constant, the operator $V\left(\boldsymbol{\lambda}\right)$ is not
changing, so that $U\left(t,t_{0}\right)=U_{\textrm{eff}(0)}\left(t,t_{0}\right)=1$
and $S\left(\omega t+\theta,t\right)=S\left(\omega t+\theta\right)$.
In that case Eq.~(\ref{eq:phi-t-solution}) becomes an exact Floquet
solution which does not have an additional slow temporal dependence:
$\left|\phi\left(\omega t+\theta,t\right)\right\rangle \equiv\left|\phi\left(\omega t+\theta\right)\right\rangle $.
By taking a set of states $\left|\phi\left(t_{0}\right)\right\rangle =\left|\alpha\right\rangle $
which form an orthonormal basis in the Hilbert space $\mathscr{H}$,
one arrives at the corresponding set of the Floquet solutions: 
\begin{equation}
\left|\phi_{\alpha}\left(\omega t+\theta\right)\right\rangle =e^{-\mathrm{i}S\left(\omega t+\theta\right)}e^{\mathrm{i}S\left(\omega t_{0}+\theta\right)}\left|\alpha\right\rangle \,.\label{eq:phi-t-solution-alpha}
\end{equation}
The solutions~(\ref{eq:phi-t-solution-alpha}) are strictly
periodic $\left|\phi_{\alpha}\left(\omega t+\theta+2\pi\right)\right\rangle =\left|\phi_{\alpha}\left(\omega t+\theta\right)\right\rangle $
and thus satisfy the Floquet theorem \cite{sambe73} with zero quasi-energies (modulus the driving energy $\hbar \omega$) for any initial state $\left|\alpha\right\rangle$.

\section{Analysis of operator $W\left(\omega t + \theta,t\right)$\label{sec:Analysis-of operator W}}

\subsection{General equations}

It is convenient to define a variable
\begin{equation}
c=c\left(\omega t + \theta \right)=\frac{\mathcal{F}\left(\omega t + \theta \right)}{\hbar\omega}
\label{eq:c}
\end{equation}
and treat   $W\left(\omega t + \theta,t\right)=\tilde{W}\left(c,t\right)$ as a function of $c$ and the slow time $t$. 
Differentiating $\tilde{W}\left(c,t\right)$ given by Eqs.~(\ref{eq:W-definition}), (\ref{eq:R-Definition}) and (\ref{eq:c}) with respect to $c$,
one arrives at the following equation    (see Appendix \ref{sec:Appendix A}): 
\begin{equation}
\frac{\partial \tilde{W}}{\partial c}=-\hbar\dot{V}+\mathrm{i}\left[V,\tilde{W}\right]\,\label{eq:W-dif equation}
\end{equation}
subject to the initial condition 
\begin{equation}
\tilde{W}\left(c,t\right)=0\,\quad\mathrm{for}\quad c=0.\label{eq:W-initial condition}
\end{equation}
Here we write the full time derivative $\dot{V}$ rather
than the partial derivative $\partial V/\partial t$, because
the  slowly changing operator $V = V\left( \boldsymbol{\lambda}\left(t\right)\right)$ does not depend on $c$. 

The solution to Eq.(\ref{eq:W-dif equation}) can be expanded in powers of $c$, giving:
\begin{align}
\tilde{W}\left(c,t\right) &= \mathrm{i}\hbar \left\{ \frac{\mathrm{i}c}{1!}\dot{V}+\frac{(\mathrm{i}c)^{2}}{2!}\left[V,\dot{V}\right] \right. \nonumber \\
 &\left.+\frac{(\mathrm{i}c)^{3}}{3!}\left[V,\left[V,\dot{V}\right]\right]+\cdots\right\}\,.
\label{eq:W-expansion}
\end{align}

\subsection{Weak driving}

Let us now consider the weak driving where Eq.~(\ref{eq:High Frequency condition general}) holds and thus it is sufficient to keep the leading terms of the expansion (\ref{eq:W-expansion}).
The first term in Eq.~(\ref{eq:W-expansion}) proportional to
$c=\mathcal{F}\left( \omega t + \theta \right)/ \hbar \omega$ 
does not have the zero frequency Fourier component 
and thus 
does not contribute to the effective Floquet Hamiltonian $W_{\textrm{eff}(0)}=W^{\left(0\right)}\left(t\right)$.
Yet, this term
provides the leading contribution to the Fourier components $W^{\left(m\right)}\left(t\right)
\approx \mathrm{i} \dot{V}f^{\left(m\right)}/m \omega$
with $m\ne0$. Thus the adiabatic condition (\ref{eq:High Frequency condition general_for_W}) 
takes the form
\begin{equation}
\left|\dot{V}_{\alpha\beta}\right|f^{\left(m\right)}\ll m\hbar\omega^{2}\,\quad \mathrm{for} \quad m\ne0\,\label{eq:adiabatic condition weak driving}
\end{equation}
in the case of the weak driving.

The second term in the expansion~(\ref{eq:W-expansion}) yields
the effective Hamiltonian for the weak driving:
\begin{equation}
W_{\mathrm{eff}(0)}\left(t\right)=W^{\left(0\right)}\left(t\right)\approx \frac{-\mathrm{i}p}{2\hbar\omega^{2}}\left[V\left(t\right),\dot{V}\left(t\right)\right]\,,\label{eq:W^(0)-weak driving-final}
\end{equation}
with 
\begin{equation}
p=\frac{1}{2\pi}\intop_{0}^{2\pi}\mathcal{F}^{2}\left(\theta^{\prime}\right)\mathrm{d}\theta^{\prime}\,.\label{eq:p}
\end{equation}
For a harmonic driving, one has $f\left(\theta\right)=\cos\theta$
and $\mathcal{F}\left(\theta\right)=\sin\theta$, giving $p=1/2$.
In that case the effective Hamiltonian (\ref{eq:W^(0)-weak driving-final})
coincides with Eq.~(38) of ref.\cite{Novicenko2017} obtained in the second order of the high frequency expansion 
of the effective Hamiltonian in the original representation. 

Generally it is not possible to obtain simple \mbox{analytical} expressions for the Floquet effective Hamiltonian $W_{\mathrm{eff}(0)}\left(t\right)=W^{\left(0\right)}\left(t\right)$, similar to Eq.~(\ref{eq:W^(0)-weak driving-final}), beyond the weak driving regime. Yet such expressions can be obtained for specific models, such as for the spin in the oscillating magnetic field. This will be considered in Sec.~\ref{subsec:W-spin}.

\subsection{$V\left(t\right)$ commutes with itself at different times} 
If $V\left(t\right)$ commutes with itself at different times, then $\left[V,\dot{V}\right]=0$, so only the first term remains in the expansion~(\ref{eq:W-expansion}), giving
\begin{equation}
W\left(\omega t + \theta,t\right)=- \mathcal{F}\left( \omega t + \theta \right) \dot{V}/ \omega \,.\label{eq:W-trivial}
\end{equation}
Since $\mathcal{F}\left( \omega t + \theta \right)$ averages to zero, 
the operator $W\left(\omega t + \theta ,t\right)$ does not have the Fourier component $W^{\left(0\right)}\left(t\right)$, so the effective Hamiltonian is equal to zero, $W^{\left(0\right)}\left(t\right)=0$. Therefore,  in order to have a non-trivial evolution giving $W^{\left(0\right)}\left(t\right)\ne0$, the operator $V\left(t\right)$ should not commute with itself at different times. For the spin in the fast oscillating magnetic field, this is the case if the direction of the magnetic field changes ($\mathbf{B}\times\dot{\mathbf{B}} \ne 0$), as we will see next.

\section{Spin in oscillating magnetic field\label{sec:Spin-in-oscillating}}

\subsection{Explicit expression for operator $W$ \label{subsec:W-spin}} 

For the spin in an oscillating magnetic field, described by the Hamiltonian~(\ref{eq:V-spin}), the operator $W$ can
be derived exactly for arbitrary strength of the periodic driving
(see the Appendix \ref{sec:Appendix B}): 
\begin{align}
W\left(\omega t+\theta,t\right) &= -\frac{g_{F} \mathcal{F}}{\omega} \frac{\left(\mathbf{B}\cdot\dot{\mathbf{B}}\right)\mathbf{B}\cdot\mathbf{F}}{B^{2}} \nonumber \\
&-\sin\left(\frac{B g_{F} \mathcal{F}}{\omega}  \right)\frac{\left[\left(\mathbf{B}\times\dot{\mathbf{B}}\right)\times\mathbf{B}\right]\cdot\mathbf{F}}{B^{3}} \nonumber\\
&-\left[\cos\left(\frac{B g_{F} \mathcal{F}}{\omega} \right)-1\right]\frac{\left(\mathbf{B}\times\dot{\mathbf{B}}\right)\cdot\mathbf{F}}{B^{2}}.
\label{eq:W-spin-solution}
\end{align}
with $\mathcal{F}=\mathcal{F}\left(\omega t + \theta \right)$ and $\mathbf{B}=\mathbf{B}(t)$.  We will use this relation to analyze the dynamics
of the spin in the oscillating magnetic field.

\subsection{Effective Hamiltonian $W_{\mathrm{eff}(0)}\left(t\right)=W^{\left(0\right)}\left(t\right)$}

The first term in Eq.~(\ref{eq:W-spin-solution}) averages to zero
and does not contribute to the effective Hamiltonian $W^{\left(0\right)}\left(t\right)$.
In what follows we will consider the harmonic driving where $f\left(\theta\right)=\cos\theta$
and hence $\mathcal{F}\left(\theta\right)=\sin\theta$. In that case
the second term of Eq.~(\ref{eq:W-spin-solution}) also averages
to zero and thus does not contribute to $W^{\left(0\right)}\left(t\right)$.
Therefore the effective Hamiltonian originates from the third term
of Eq.~(\ref{eq:W-spin-solution}) and is given by 

\begin{equation}
W^{\left(0\right)}\left(t\right)=\frac{\left[1-\mathcal{J}_{0}\left(g_{F}B/ \omega \right)\right]}{B^{2}}\mathbf{F}\cdot\left(\mathbf{B}\times\dot{\mathbf{B}}\right)\,,
\label{eq:W^(0)_harmonic-Spin-Result-explicit}
\end{equation}
where $\mathcal{J}_{0}\left(a\right)=\frac{1}{2\pi}\intop_{0}^{2\pi}e^{\mathrm{i}a\sin\theta}\mathrm{d}\theta$
is the zero-order Bessel function, $B=\left|\mathbf{B}\right|$, and the time dependence of $\mathbf{B}=\mathbf{B}\left(t\right)$
is kept implicit. For $g_{F}B/\omega\ll1$, Eq.~(\ref{eq:W^(0)_harmonic-Spin-Result-explicit})
reduces to the previous result \cite{Novicenko2017} applicable for
the weak driving: 
\begin{equation}
W^{\left(0\right)}\left(t\right)\approx\frac{g_{F}^{2}}{4\omega^{2}}\mathbf{F}\cdot\left(\mathbf{B}\times\dot{\mathbf{B}}\right)\,.\label{eq:W^(0)_harmonic-Spin-Result-small B}
\end{equation}

Introducing a unit vector along the magnetic field ${\mathbf{b}=\mathbf{B}/B}$,
one has $\mathbf{b}\times\dot{\mathbf{b}}=\Omega\mathbf{n}$,
where $\Omega$ can be interpreted as a frequency of the magnetic
field rotation around an instantaneous rotation axis pointing along a
unit vector $\mathbf{n}$. With these notations Eq.~(\ref{eq:W^(0)_harmonic-Spin-Result-explicit})
takes the form 
\begin{equation}
W^{\left(0\right)}\left(t\right)=\Omega\left[1-\mathcal{J}_{0}\left(g_{F}B/ \omega \right)\right]\mathbf{F}\cdot\mathbf{n}\,.\label{eq:W^(0)_harmonic-Spin-Result-explicit-alternative}
\end{equation}

\subsection{Adiabatic evolution}

\begin{figure}[h!]
\centering\includegraphics[width=0.98\columnwidth]{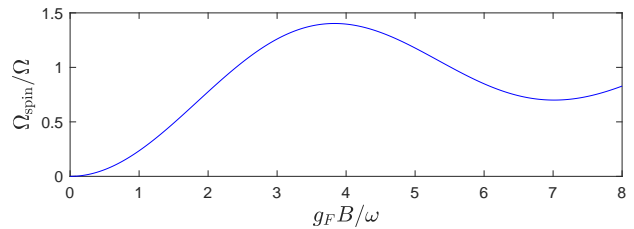}
\caption{\label{fig:bessel} Dependence of $\Omega_{\mathrm{spin}}/ \Omega$ on the ratio $g_F B/\omega$.}
\end{figure}

If the magnetic field rotates with a frequency $\Omega$
in a plane perpendicular to a fixed axis $\mathbf{n}$, the effective
Hamiltonian given by Eqs.~(\ref{eq:W^(0)_harmonic-Spin-Result-explicit})
or (\ref{eq:W^(0)_harmonic-Spin-Result-explicit-alternative}) describes
the spin rotation around $\mathbf{n}$ with a frequency 
\begin{equation}
\Omega_{\mathrm{spin}}=\Omega\left[1-\mathcal{J}_{0}\left(a\right)\right],\;\mathrm{with}\; a=g_{F}B/\omega.
\label{eq:Omega_spin}
\end{equation}
For $a\ll1$, the frequency ${\Omega_{\mathrm{spin}}\approx\Omega a^2/4}$
is much smaller than the rotation frequency $\Omega$ of the magnetic
field. With an increase of $a$, the frequency $\Omega_{\mathrm{spin}}$
increases. In particular, for $a\approx2.4$ the Bessel function $\mathcal{J}_{0}\left(a\right)$
becomes zero, and the spin rotates with the same frequency as the
magnetic field: $\Omega_{\mathrm{spin}}=\Omega$ (see Fig.~\ref{fig:bessel}). By further increasing
$a$, one arrives at a regime where $\mathcal{J}_{0}\left(a\right)<0$,
and the frequency $\Omega_{\mathrm{spin}}$ exceeds $\Omega$. The
maximum frequency of the spin rotation $\Omega_{\mathrm{spin}}=1.36\Omega$
is achieved for $a\approx3.83$ where the Bessel function $\mathcal{J}_{0}\left(a\right)$
has its minimum, as illustrated in Fig.~\ref{fig:bessel}.

If the direction $\mathbf{n}$ of the rotation axis is
changing, the Hamiltonian $ W^{\left(0\right)}\left(t\right)$
does not commute with itself at different times, and the time ordering
is needed in the effective evolution operator~(\ref{U_eff-definition}). 
Therefore the effective evolution of the spin is associated with non-Abelian
(non-commuting) geometric phases. In the present situation, the magnetic
field $\mathbf{B}\left(t\right)$ plays the role of the slowly varying
$\boldsymbol{\lambda}\left(t\right)$ featured in Sec.~\ref{subsec:Non-Abelian-geometric-phases},
so the effective Hamiltonian~(\ref{eq:W^(0)_harmonic-Spin-Result-explicit})
can be represented in terms of the non-Abelian vector potential $\mathbf{A}^{\left(0\right)}\left(t\right)$:
\begin{equation}
W^{\left(0\right)}\left(t\right)=\dot{\mathbf{B}} \cdot \mathbf{A}^{\left(0\right)} \,,
\label{eq:W^(0)_in terms of_A^(0) spin}
\end{equation}
with
\begin{equation}
\mathbf{A}^{\left(0\right)}=\frac{\left[1-\mathcal{J}_{0}\left(g_F B/\omega\right)\right]}{B^{2}}\left(\mathbf{B}\times\mathbf{F}\right).
\label{eq:A^(0) spin}
\end{equation}
The evolution operator $U_{\textrm{eff}(0)}\left(t,t_{0}\right)$
is then given by Eq.~(\ref{U_eff-in-terms-of-vector potential})
with $\boldsymbol{\lambda}\left(t\right)$ replaced by $\mathbf{B}\left(t\right)$.

In particular, one can perform a cyclic anti-clockwise rotation of the
magnetic field $\mathbf{B}$ in a plane orthogonal to a fixed unit
vector $\mathbf{n}$ without changing the modulus $B$.  
Using Eq.~(\ref{eq:W^(0)_in terms of_A^(0) spin}), the evolution
operator~(\ref{U_eff-in-terms-of-vector potential}) then reads 
\begin{equation}
U_{\textrm{eff}(0)}^{\left(\mathbf{n}\right)}=\exp\left[-\frac{\mathrm{i}}{\hbar}\gamma\mathbf{F}\cdot\mathbf{n}\right], \; \gamma=2\pi\left[1-\mathcal{J}_{0}\left(g_F B/\omega\right)\right]\,.
\label{U_eff-spin-n-axis}
\end{equation}
The operator $\gamma\mathbf{F}\cdot\mathbf{n}/\hbar$ provides the geometric
phase $\gamma m_{F}$ for the spin with the projection $m_{F}$ along
the rotation axis $\mathbf{n}$. For weak driving ($g_{F}B/\omega\ll 1$)
the geometric phase $\gamma m_{F}$ is much smaller than unity,
and the magnetic field $\mathbf{B}$ has to complete many rotation
cycles to accumulate a considerable geometric phase \cite{Novicenko2017}.
On the other hand, if $g_{F}B$ is comparable with $\omega$, a sizable
geometric phase is acquired during a single cycle. Therefore two consecutive
rotations along non-parallel axes $\mathbf{n}$ and $\mathbf{n}^{\prime}$
do not commute, $\left[U_{\textrm{eff}(0)}^{\left(\mathbf{n}\right)},U_{\textrm{eff}(0)}^{\left(\mathbf{n}^{\prime}\right)}\right]\ne0$,
and the corresponding geometric phases $\gamma\mathbf{F}\cdot\mathbf{n}/\hbar$ and $\gamma\mathbf{F}\cdot\mathbf{n^{\prime}}/\hbar$ are non-Abelian. 

Previously, Berry analyzed a spin that adiabatically follows a slowly changing magnetic field  \cite{Berry:1984}. After the magnetic field vector completes a closed loop trajectory and returns to its initial value, the state-vector for the spin acquires a 
geometric (Berry) phase factor. 
Such a phase factor belongs to the Abelian group $U(1)$. For the periodically driven spin considered here, the adiabatic
evolution of the state-vector in the degenerate Floquet manifold is
described by the geometric phase operator $U_{\textrm{eff}(0)}\left(t,t_{0}\right)=\exp\left(\mathrm{i}\Gamma\right)$
belonging to the non-Abelian group $SU(2)$. Therefore the periodic
driving enriches the system. Note that in the present situation the non-Abelian Floquet geometric
phases appear by adiabatically eliminating other Floquet bands rather
than by eliminating other states of the physical Hilbert space $\mathscr{H}$,
as it is the case for non-driven systems \cite{Wilczek:1984,Moody1986,Zee1988}.

\subsection{Adiabatic condition and micromotion}

The previous analysis of the spin in the oscillating magnetic field
\cite{Novicenko2017} relies on the high frequency assumption for the magnetic field amplitude, $g_{F}f_{F}B\left(t\right)\ll\omega$,
and for its changes.
Now we only require that $\mathbf{B}\left(t\right)$
changes sufficiently slowly, so that the adiabatic condition~(\ref{eq:High Frequency condition general_for_W})
holds for $W$ given by Eq.~(\ref{eq:W-spin-solution}). The general expression for the  adiabatic condition given by Eqs.~(\ref{eq:High Frequency condition general_for_W}) and (\ref{eq:W-spin-solution}) is quite cumbersome. Yet if 
\begin{equation}
g_{F}f_{F}\left|\dot{\mathbf{B}}\left(t\right)\right|/\omega\ll\omega\,,\label{eq:Adiabatic condition--weak driving__Spin in B field}
\end{equation}
the adiabatic condition is fulfilled for any strength of the magnetic field. In other words, if (\ref{eq:Adiabatic condition--weak driving__Spin in B field}) holds then the adiabatic condition~(\ref{eq:High Frequency condition general_for_W}) is fulfilled but not vice versa.

The micromotion is described by the Hermitian operator $S\left(\omega t+\theta,t\right)$
entering the full evolution operator in Eq.~(\ref{eq:phi-t-solution}).
Using Eqs.~(\ref{eq:V-spin}) and (\ref{eq:S}), the micromotion operator reads for the spin in the magnetic field
\begin{equation}
S\left(\omega t+\theta,t\right)=\frac{g_{F}}{\hbar\omega}\mathbf{F}\cdot\mathbf{B}\left(t\right)\sin\left(\omega t+\theta\right)\,.\label{eq:S-Spin-in-magn-field}
\end{equation}
The micromotion increases with increasing the magnetic field strength
and becomes substantial when the high frequency condition ($g_{F}f_{F}B\left(t\right)\ll\omega$)
no longer holds. Yet, if the magnetic field is ramped up and down
slowly, the micromotion does affect the overall evolution of the system, as it was generally shown in Eq.~(\ref{eq:phi-t-solution-slow ramping})
and will be discussed in more details next.

\subsection{Specific sequence of the magnetic field\label{subsec:Specific-sequence-of} }

The non-Abelian geometric phases can be measured using, for example,
the following sequence for the oscillating magnetic field. At the
initial time $t=t_{0}$ the magnetic field is zero ($\mathbf{B}\left(t_{0}\right)\rightarrow0$)
and is ramped up smoothly afterwards, so there is no contribution
by the micromotion due to switching on the magnetic field: $S\left(\theta+\omega t_{0},t_{0}\right)=0$.
For $t_{0}<t<t_{1}$ the amplitude of magnetic field increases from
zero to a steady state value $\mathbf{B}\left(t\right)=B_{0}\mathbf{e}_{z}$
without changing its direction. Therefore the effective Hamiltonian
$W_{\textrm{eff}(0)}=W^{(0)}$ is zero in this stage, giving $U_{\textrm{eff}(0)}\left(t_{1},t_{0}\right)=1$.
During the subsequent evolution at $t_{1}<t<t_{2}$ the magnetic field
changes its direction while keeping constant the modulus $B$. For
example, the magnetic field can undergo two consecutive cycles of
rotation, around the $y$ and $x$ axes, described by non-commuting
unitary operators $U_{\textrm{eff}(0)}^{\left(\mathbf{e}_{y}\right)}$
and $U_{\textrm{eff}(0)}^{\left(\mathbf{e}_{x}\right)}$ given by Eq.~(\ref{U_eff-spin-n-axis}),
and the effective evolution operator reads $U_{\textrm{eff}(0)}\left(t_{2},t_{1}\right)=U_{\textrm{eff}(0)}^{\left(\mathbf{e}_{x}\right)}U_{\textrm{eff}(0)}^{\left(\mathbf{e}_{y}\right)}$.
Note that for $t_{1}<t<t_{0}$ the magnetic field strength can be
considerable, and the high frequency condition $g_{F}f_{F}B\left(t\right)\ll\omega$
does not necessarily hold. Therefore the evolution operator $U_{\textrm{eff}(0)}\left(t_{2},t_{1}\right)$
can significantly alter the state-vector of the system. Finally, for
$t_{2}<t<t_{3}$ the magnetic field is ramped down to zero without
changing its direction, giving $U_{\textrm{eff}(0)}\left(t_{3},t_{2}\right)=1$
and $S\left(\theta+\omega t_{3},t_{3}\right)=0$. In this way, the
full evolution of the state-vector from $t=t_{0}$ to $t=t_{3}$
is given by Eq.~(\ref{eq:phi-t-solution-slow ramping}) with $t$ replaced by $t_3$. The evolution of the state-vector is thus described exclusively by the operator $U_{\textrm{eff}(0)}\left(t_{2},t_{1}\right)$
which is does not depend on the details of the ramping up and down
of the magnetic field.

\begin{figure}[h]
\begin{centering}
\includegraphics[scale=0.5]{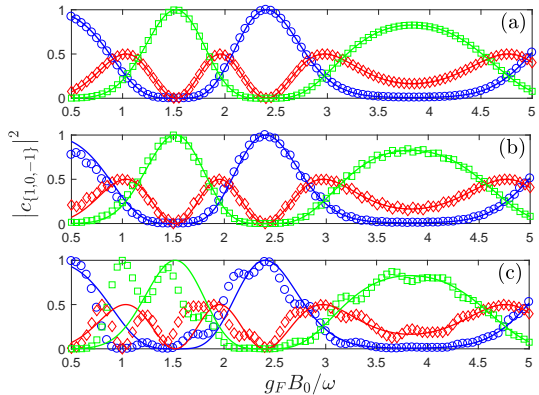}
\par\end{centering}
\caption{\label{fig:2} Comparison of the analytical (solid lines) and exactly
calculated (symbols) evolution for the spin $1$ (${f_{F}=1}$). Here
$\left|c_{m_{F}^{\prime}}\right|^{2}$ are the probabilities for
the spin projection along the $z$ axis to be $m_{F}^{\prime}=0,\pm1$
at the final time $t=t_{3}$. At the initial time $t=t_{0}$ the spin
is along the $z$ axis ($m_{F}=1$). Blue circles, red diamonds and
green squares correspond to $\left|c_{1}\right|^{2}$, $\left|c_{0}\right|^{2}$
and $\left|c_{-1}\right|^{2}$, respectively. The effective evolution
described by the operator $U_{\textrm{eff}(0)}\left(t_{2},t_{1}\right)=U_{\textrm{eff}(0)}^{\left(\mathbf{e}_{y}\right)}$
provides the following probabilities $\left|c_{\left\{ 1,0,-1\right\} }\right|^{2}=\left\{ \cos^{4}\left(\gamma/2\right),\,\sin^{2}\left(\gamma\right)/2\,,\sin^{4}\left(\gamma/2\right)\right\} $ plotted by the solid lines,
with $\gamma$ being defined in Eq.~(\ref{U_eff-spin-n-axis}). The ratio $\rho=\Omega g_{F}B_{0}/\omega^{2}$ equals to $\rho=0.1$, $\rho=0.3$ and $\rho=1$
in (a), (b) and (c) panels, respectively.}
\end{figure}
In Fig \ref{fig:2} we have checked the validity of the description
of the evolution of the system in terms of the effective evolution
operator $U_{\textrm{eff}(0)}\left(t_{2},t_{1}\right)$. We have plotted the
exact and effective evolution of the spin $1$ system ($f_{F}=1$) from $t=t_{0}$
to $t=t_{3}$ for different angular frequencies $\Omega$ of rotation of the 
magnetic field direction during the second stage where $t_{1}<t<t_{2}$.
We have considered the case where the magnetic field
completes a single cycle of rotation around the $y$ axis from $t=t_1$ to $t=t_2$,
so that $U_{\textrm{eff}(0)}\left(t_{2},t_{1}\right)=U_{\textrm{eff}(0)}^{\left(\mathbf{e}_{y}\right)}$.
The angular frequency $\Omega$ is chosen such that the ratio $\rho=\Omega g_{F}B_{0}/\omega^{2}$
is not changing in the same plot. The exact and analytical results agree well if the rotation is sufficiently
slow, $\rho\ll1$, and thus the adiabatic condition (\ref{eq:Adiabatic condition--weak driving__Spin in B field})
holds. 

\section{Concluding remarks\label{sec:Concluding-remarks}}

We have considered the evolution of a periodically driven quantum
system governed by the Hamiltonian $H\left(\omega t+\theta,t\right)$
which is the product of a slowly varying Hermitian operator $V\left(\boldsymbol{\lambda}\left(t\right)\right)$
and a fast oscillating periodic function $f\left(\omega t+\theta\right)$
with zero average. The analysis does not rely on the high frequency
approximation \cite{Novicenko2017} for the original Hamiltonian $H\left(\omega t+\theta,t\right)$,
so the driving frequency $\omega$ can be both larger or smaller than
 the matrix elements of the slowly changing operator $V\left(\boldsymbol{\lambda}\left(t\right)\right)/\hbar$.
 We have shown that
the adiabatic evolution of the system within a degenerate Floquet band
is accompanied by the non-Abelian (non-commuting) Floquet geometric
phases which can  be significant even after completing
a single cycle of the slow variable $\boldsymbol{\lambda}=
\boldsymbol{\lambda}\left(t\right)$.
On the other hand, for the weak driving
the geometric phases acquired during a cyclic evolution of the slow
variable are small, and the slow variable should complete many cycles
to accumulate considerable geometric phases~\cite{Novicenko2017}.

Without the periodic driving $f\left(\omega t+\theta\right)$ the
spin adiabatically follows the slowly changing magnetic field, and
the adiabatic elimination of other spin states provides the Berry
phase  factor \cite{Berry:1984} which belongs to the Abelian group $U(1)$. The periodic driving enriches
the system, and the non-Abelian geometric phases appear by adiabatically
eliminating other Floquet bands rather than by eliminating other states
of the physical Hilbert space $\mathscr{H}$, as it is the case for
non-driven systems \cite{Wilczek:1984,Moody1986,Zee1988}. In the
latter non-driven systems non-Abelian geometric phases can be formed if there
is a manifold of degenerate physical states well separated from other
states,  such as a pair of degenerate dark states in the tripod atom-light
coupling scheme \cite{Unanyan99PRA,Theuer1990OE,Unanyan2004,Ruseckas2005,Leroux2018NCommun} 
or a pair of degenerate spin-up and spin-down states emerging in the nuclear quadrupole
resonance \cite{Zee1988} and for diatomic molecules \cite{Moody1986,Bohm92JMP}.
Note also that the previous studies of the Floquet adiabatic perturbation theory \cite{Weinberg17PR} dealt with the 
non-degenerate Floquet states, so the emerging geometric phases are Abelian. 

A distinctive feature of the present periodically driven system is that the Floquet
eigen-energies are fully degenerate within individual Floquet bands
even if the eigen-energies of the slowly varying part of the original
Hamiltonian $V\left(\boldsymbol{\lambda}\left(t\right)\right)$ are not degenerate. Therefore the non-Abelian
geometric phases emerge in a very straightforward way, and no degeneracy
of the physical states is needed. Furthermore, the individual Floquet
bands are characterized by zero energy (modulus the driving
frequency), so there are no unwanted dynamical phases accompanying
the non-Abelian geometric phases during the adiabatic evolution of
the system within individual Floquet bands.
This is because the dynamical phases average to zero over an oscillation
period.

The adiabatic evolution of periodically driven systems is generally
accompanied by micromotion. Yet the effects of the micromotion can
be avoided if the periodic driving is ramped up slowly at the initial
stage and subsequently ramped down slowly at the final stage.
The dynamics of the state-vector given by Eq.~(\ref{eq:phi-t-solution-slow ramping})
is then represented exclusively by the 
operator $U_{\textrm{eff}(0)}\left(t,t_{0}\right)$ describing the non-Abelian geometric phases  emerging
for the 
adiabatic evolution of the system within a degenerate Floquet band. The geometric phases
are determined by the trajectory of the slowly varying parameters
$\boldsymbol{\lambda}$, rather than by the speed at which these parameters
change. This can be used for precisely controlling the evolution of
quantum systems, in particular for realization of fault-tolerant quantum gates~\cite{Danilin_2018}.

\section*{Acknowledgments}

We acknowledge helpful discussions with Egidijus Anisimovas, Julius
Ruseckas and Thomas Gajdosik. This Research was funded by European Social Fund (Project No. 09.3.3--LMT--K--712--02--0065). 

\appendix 
\section{Equation for operator $W$ \label{sec:Appendix A}}

Differentiating $W\left(\omega t + \theta,t\right)=\tilde{W}\left(c,t\right)$ given by Eqs.~(\ref{eq:W-definition}), (\ref{eq:R-Definition}) and (\ref{eq:c}) with respect to $c$ for fixed slow time $t$,
one has
\begin{equation}
\frac{\partial \tilde{W}}{\partial c}=\hbar VR^{\dagger}\frac{\partial}{\partial t}R- \hbar R^{\dagger}\frac{\partial}{\partial t}\left(RV\right)\,.\label{eq:W-derivative}
\end{equation}
Since $\hbar R^{\dagger}\frac{\partial}{\partial t}\left(RV\right)=\hbar  \dot{V}+\mathrm{i}\tilde{W}V$,
Eq.~(\ref{eq:W-derivative}) yields the differential equation for
$\tilde{W}$
\begin{equation}
\frac{\partial \tilde{W}}{\partial c}=-\hbar\dot{V}+\mathrm{i}\left[V,\tilde{W}\right]\,\label{eq:W-dif equation-1}
\end{equation}
subject to the initial condition: $\tilde{W}\left(c,t\right)=0$ for $c=0$.

\section{Operator $W$ for spin in oscillating magnetic field \label{sec:Appendix B}}

Let us now find a solution to Eq.~(\ref{eq:W-dif equation-1}) for a spin in an oscillating magnetic field. In that case
the slowly varying part of the Hamiltonian is given by Eq.~(\ref{eq:V-spin}).
We are looking for a solution in the form 
\begin{equation}
\tilde{W}\left(c,t\right)=\mathbf{F}\cdot\mathbf{X}\,.\label{eq:W-spin-Ansatz}
\end{equation}
Substituting Eq.~(\ref{eq:W-spin-Ansatz}) into (\ref{eq:W-dif equation-1}),
one arrives at the following equation for the vector $\mathbf{X}=\mathbf{X}\left(c,t\right)$
\begin{equation}
\frac{\partial\mathbf{X}}{\partial c}=-\hbar g_{F}\dot{\mathbf{B}}-\hbar g_{F}\mathbf{B}\times\mathbf{X}\,,\label{eq:X-dif equation}
\end{equation}
with the initial condition $\mathbf{X}\left(c,t\right)=0$ for $c=0$.
A solution to this equation is:
\begin{align}
\mathbf{X}\left(c,t\right) &= - c\hbar g_{F}\frac{\left(\mathbf{B}\cdot\dot{\mathbf{B}}\right)\mathbf{B}}{B^{2}} \nonumber \\
&-\sin\left(c\hbar g_{F}B\right)\frac{\left(\mathbf{B}\times\dot{\mathbf{B}}\right)\times\mathbf{B}}{ B^{3}} \nonumber \\
&-\left[\cos\left(c\hbar g_{F}B\right)-1\right]\frac{\mathbf{B}\times\dot{\mathbf{B}}}{ B^{2}}.
\label{eq:X-Ansatz-result-1}
\end{align}
Equations~(\ref{eq:W-spin-Ansatz}), (\ref{eq:X-Ansatz-result-1}) and (\ref{eq:c})
provide the explicit expression for $\tilde{W}\left(c,t\right)=W\left(\omega t + \theta,t\right)$ presented by Eq.~(\ref{eq:W-spin-solution}) in the main text.

\bibliography{Floquet-extra-space2018}

\begin{thebibliography}{79}%
\makeatletter
\providecommand \@ifxundefined [1]{%
 \@ifx{#1\undefined}
}%
\providecommand \@ifnum [1]{%
 \ifnum #1\expandafter \@firstoftwo
 \else \expandafter \@secondoftwo
 \fi
}%
\providecommand \@ifx [1]{%
 \ifx #1\expandafter \@firstoftwo
 \else \expandafter \@secondoftwo
 \fi
}%
\providecommand \natexlab [1]{#1}%
\providecommand \enquote  [1]{``#1''}%
\providecommand \bibnamefont  [1]{#1}%
\providecommand \bibfnamefont [1]{#1}%
\providecommand \citenamefont [1]{#1}%
\providecommand \href@noop [0]{\@secondoftwo}%
\providecommand \href [0]{\begingroup \@sanitize@url \@href}%
\providecommand \@href[1]{\@@startlink{#1}\@@href}%
\providecommand \@@href[1]{\endgroup#1\@@endlink}%
\providecommand \@sanitize@url [0]{\catcode `\\12\catcode `\$12\catcode
  `\&12\catcode `\#12\catcode `\^12\catcode `\_12\catcode `\%12\relax}%
\providecommand \@@startlink[1]{}%
\providecommand \@@endlink[0]{}%
\providecommand \url  [0]{\begingroup\@sanitize@url \@url }%
\providecommand \@url [1]{\endgroup\@href {#1}{\urlprefix }}%
\providecommand \urlprefix  [0]{URL }%
\providecommand \Eprint [0]{\href }%
\providecommand \doibase [0]{http://dx.doi.org/}%
\providecommand \selectlanguage [0]{\@gobble}%
\providecommand \bibinfo  [0]{\@secondoftwo}%
\providecommand \bibfield  [0]{\@secondoftwo}%
\providecommand \translation [1]{[#1]}%
\providecommand \BibitemOpen [0]{}%
\providecommand \bibitemStop [0]{}%
\providecommand \bibitemNoStop [0]{.\EOS\space}%
\providecommand \EOS [0]{\spacefactor3000\relax}%
\providecommand \BibitemShut  [1]{\csname bibitem#1\endcsname}%
\let\auto@bib@innerbib\@empty
\bibitem [{\citenamefont {Oka}\ and\ \citenamefont {Aoki}(2009)}]{Oka2009}%
  \BibitemOpen
  \bibfield  {author} {\bibinfo {author} {\bibfnamefont {T.}~\bibnamefont
  {Oka}}\ and\ \bibinfo {author} {\bibfnamefont {H.}~\bibnamefont {Aoki}},\
  }\href {\doibase 10.1103/PhysRevB.79.081406} {\bibfield  {journal} {\bibinfo
  {journal} {Phys. Rev. B}\ }\textbf {\bibinfo {volume} {79}},\ \bibinfo
  {pages} {081406(R)} (\bibinfo {year} {2009})}\BibitemShut {NoStop}%
\bibitem [{\citenamefont {Kitagawa}\ \emph {et~al.}(2010)\citenamefont
  {Kitagawa}, \citenamefont {Berg}, \citenamefont {Rudner},\ and\ \citenamefont
  {Demler}}]{Kitagawa2010}%
  \BibitemOpen
  \bibfield  {author} {\bibinfo {author} {\bibfnamefont {T.}~\bibnamefont
  {Kitagawa}}, \bibinfo {author} {\bibfnamefont {E.}~\bibnamefont {Berg}},
  \bibinfo {author} {\bibfnamefont {M.}~\bibnamefont {Rudner}}, \ and\ \bibinfo
  {author} {\bibfnamefont {E.}~\bibnamefont {Demler}},\ }\href {\doibase
  10.1103/PhysRevB.82.235114} {\bibfield  {journal} {\bibinfo  {journal} {Phys.
  Rev. B}\ }\textbf {\bibinfo {volume} {82}},\ \bibinfo {pages} {235114}
  (\bibinfo {year} {2010})}\BibitemShut {NoStop}%
\bibitem [{\citenamefont {Kitagawa}\ \emph {et~al.}(2011)\citenamefont
  {Kitagawa}, \citenamefont {Oka}, \citenamefont {Brataas}, \citenamefont
  {Fu},\ and\ \citenamefont {Demler}}]{Kitagawa2011}%
  \BibitemOpen
  \bibfield  {author} {\bibinfo {author} {\bibfnamefont {T.}~\bibnamefont
  {Kitagawa}}, \bibinfo {author} {\bibfnamefont {T.}~\bibnamefont {Oka}},
  \bibinfo {author} {\bibfnamefont {A.}~\bibnamefont {Brataas}}, \bibinfo
  {author} {\bibfnamefont {L.}~\bibnamefont {Fu}}, \ and\ \bibinfo {author}
  {\bibfnamefont {E.}~\bibnamefont {Demler}},\ }\href {\doibase
  10.1103/PhysRevB.84.235108} {\bibfield  {journal} {\bibinfo  {journal} {Phys.
  Rev. B}\ }\textbf {\bibinfo {volume} {84}},\ \bibinfo {pages} {235108}
  (\bibinfo {year} {2011})}\BibitemShut {NoStop}%
\bibitem [{\citenamefont {Lindner}\ \emph {et~al.}(2011)\citenamefont
  {Lindner}, \citenamefont {Refael},\ and\ \citenamefont
  {Galitski}}]{Galitski2011NP}%
  \BibitemOpen
  \bibfield  {author} {\bibinfo {author} {\bibfnamefont {N.~H.}\ \bibnamefont
  {Lindner}}, \bibinfo {author} {\bibfnamefont {G.}~\bibnamefont {Refael}}, \
  and\ \bibinfo {author} {\bibfnamefont {V.}~\bibnamefont {Galitski}},\ }\href
  {\doibase 10.1038/nphys1926} {\bibfield  {journal} {\bibinfo  {journal} {Nat.
  Phys.}\ }\textbf {\bibinfo {volume} {7}},\ \bibinfo {pages} {490} (\bibinfo
  {year} {2011})}\BibitemShut {NoStop}%
\bibitem [{\citenamefont {Kolovsky}(2011)}]{kolovsky11}%
  \BibitemOpen
  \bibfield  {author} {\bibinfo {author} {\bibfnamefont {A.~R.}\ \bibnamefont
  {Kolovsky}},\ }\href {\doibase 10.1209/0295-5075/93/20003} {\bibfield
  {journal} {\bibinfo  {journal} {Europhys. Lett.}\ }\textbf {\bibinfo {volume}
  {93}},\ \bibinfo {pages} {20003} (\bibinfo {year} {2011})}\BibitemShut
  {NoStop}%
\bibitem [{\citenamefont {{Creffield}}\ and\ \citenamefont
  {{Sols}}(2013)}]{creffield13comment}%
  \BibitemOpen
  \bibfield  {author} {\bibinfo {author} {\bibfnamefont {C.~E.}\ \bibnamefont
  {{Creffield}}}\ and\ \bibinfo {author} {\bibfnamefont {F.}~\bibnamefont
  {{Sols}}},\ }\href {\doibase 10.1209/0295-5075/101/40001} {\bibfield
  {journal} {\bibinfo  {journal} {Europhys. Lett.}\ }\textbf {\bibinfo {volume}
  {101}},\ \bibinfo {pages} {40001} (\bibinfo {year} {2013})}\BibitemShut
  {NoStop}%
\bibitem [{\citenamefont {Rudner}\ \emph {et~al.}(2013)\citenamefont {Rudner},
  \citenamefont {Lindner}, \citenamefont {Berg},\ and\ \citenamefont
  {Levin}}]{Rudner2013}%
  \BibitemOpen
  \bibfield  {author} {\bibinfo {author} {\bibfnamefont {M.~S.}\ \bibnamefont
  {Rudner}}, \bibinfo {author} {\bibfnamefont {N.~H.}\ \bibnamefont {Lindner}},
  \bibinfo {author} {\bibfnamefont {E.}~\bibnamefont {Berg}}, \ and\ \bibinfo
  {author} {\bibfnamefont {M.}~\bibnamefont {Levin}},\ }\href {\doibase
  10.1103/PhysRevX.3.031005} {\bibfield  {journal} {\bibinfo  {journal} {Phys.
  Rev. X}\ }\textbf {\bibinfo {volume} {3}},\ \bibinfo {pages} {031005}
  (\bibinfo {year} {2013})}\BibitemShut {NoStop}%
\bibitem [{\citenamefont {Nathan}\ and\ \citenamefont
  {Rudner}(2015)}]{Nathan15NJP}%
  \BibitemOpen
  \bibfield  {author} {\bibinfo {author} {\bibfnamefont {F.}~\bibnamefont
  {Nathan}}\ and\ \bibinfo {author} {\bibfnamefont {M.~S.}\ \bibnamefont
  {Rudner}},\ }\href@noop {} {\bibfield  {journal} {\bibinfo  {journal} {New J.
  Phys.}\ }\textbf {\bibinfo {volume} {17}},\ \bibinfo {pages} {125014}
  (\bibinfo {year} {2015})}\BibitemShut {NoStop}%
\bibitem [{\citenamefont {Weinberg}\ \emph {et~al.}(2017)\citenamefont
  {Weinberg}, \citenamefont {Bukov}, \citenamefont {D'Alessio}, \citenamefont
  {Polkovnikov}, \citenamefont {Vajna},\ and\ \citenamefont
  {Kolodrubetz}}]{Weinberg17PR}%
  \BibitemOpen
  \bibfield  {author} {\bibinfo {author} {\bibfnamefont {P.}~\bibnamefont
  {Weinberg}}, \bibinfo {author} {\bibfnamefont {M.}~\bibnamefont {Bukov}},
  \bibinfo {author} {\bibfnamefont {L.}~\bibnamefont {D'Alessio}}, \bibinfo
  {author} {\bibfnamefont {A.}~\bibnamefont {Polkovnikov}}, \bibinfo {author}
  {\bibfnamefont {S.}~\bibnamefont {Vajna}}, \ and\ \bibinfo {author}
  {\bibfnamefont {M.}~\bibnamefont {Kolodrubetz}},\ }\href@noop {} {\bibfield
  {journal} {\bibinfo  {journal} {Phys. Rep.}\ }\textbf {\bibinfo {volume}
  {688}},\ \bibinfo {pages} {1} (\bibinfo {year} {2017})},\ \Eprint
  {http://arxiv.org/abs/1606.02229} {arXiv:1606.02229} \BibitemShut {NoStop}%
\bibitem [{\citenamefont {Kolodrubetz}\ \emph {et~al.}(2018)\citenamefont
  {Kolodrubetz}, \citenamefont {Nathan}, \citenamefont {Gazit}, \citenamefont
  {Morimoto},\ and\ \citenamefont {Moore}}]{Kolodrubetz18PRL}%
  \BibitemOpen
  \bibfield  {author} {\bibinfo {author} {\bibfnamefont {M.~H.}\ \bibnamefont
  {Kolodrubetz}}, \bibinfo {author} {\bibfnamefont {F.}~\bibnamefont {Nathan}},
  \bibinfo {author} {\bibfnamefont {S.}~\bibnamefont {Gazit}}, \bibinfo
  {author} {\bibfnamefont {T.}~\bibnamefont {Morimoto}}, \ and\ \bibinfo
  {author} {\bibfnamefont {J.~E.}\ \bibnamefont {Moore}},\ }\href {\doibase
  10.1103/PhysRevLett.120.150601} {\bibfield  {journal} {\bibinfo  {journal}
  {Phys. Rev. Lett.}\ }\textbf {\bibinfo {volume} {120}},\ \bibinfo {pages}
  {150601} (\bibinfo {year} {2018})}\BibitemShut {NoStop}%
\bibitem [{\citenamefont {Sacha}\ and\ \citenamefont
  {Zakrzewski}(2018)}]{Sacha18RMP}%
  \BibitemOpen
  \bibfield  {author} {\bibinfo {author} {\bibfnamefont {K.}~\bibnamefont
  {Sacha}}\ and\ \bibinfo {author} {\bibfnamefont {J.}~\bibnamefont
  {Zakrzewski}},\ }\href@noop {} {\bibfield  {journal} {\bibinfo  {journal}
  {Rep. Progr. Phys.}\ }\textbf {\bibinfo {volume} {81}},\ \bibinfo {pages}
  {016401} (\bibinfo {year} {2018})}\BibitemShut {NoStop}%
\bibitem [{\citenamefont {Fregoso}\ \emph {et~al.}(2013)\citenamefont
  {Fregoso}, \citenamefont {Wang}, \citenamefont {Gedik},\ and\ \citenamefont
  {Galitski}}]{Galitski13PRB}%
  \BibitemOpen
  \bibfield  {author} {\bibinfo {author} {\bibfnamefont {B.~M.}\ \bibnamefont
  {Fregoso}}, \bibinfo {author} {\bibfnamefont {Y.~H.}\ \bibnamefont {Wang}},
  \bibinfo {author} {\bibfnamefont {N.}~\bibnamefont {Gedik}}, \ and\ \bibinfo
  {author} {\bibfnamefont {V.}~\bibnamefont {Galitski}},\ }\href {\doibase
  10.1103/PhysRevB.88.155129} {\bibfield  {journal} {\bibinfo  {journal} {Phys.
  Rev. B}\ }\textbf {\bibinfo {volume} {88}},\ \bibinfo {pages} {155129}
  (\bibinfo {year} {2013})}\BibitemShut {NoStop}%
\bibitem [{\citenamefont {{Tong}}\ \emph {et~al.}(2013)\citenamefont {{Tong}},
  \citenamefont {{An}}, \citenamefont {{Gong}}, \citenamefont {{Luo}},\ and\
  \citenamefont {{Oh}}}]{tong13majorana}%
  \BibitemOpen
  \bibfield  {author} {\bibinfo {author} {\bibfnamefont {Q.-J.}\ \bibnamefont
  {{Tong}}}, \bibinfo {author} {\bibfnamefont {J.-H.}\ \bibnamefont {{An}}},
  \bibinfo {author} {\bibfnamefont {J.}~\bibnamefont {{Gong}}}, \bibinfo
  {author} {\bibfnamefont {H.-G.}\ \bibnamefont {{Luo}}}, \ and\ \bibinfo
  {author} {\bibfnamefont {C.~H.}\ \bibnamefont {{Oh}}},\ }\href {\doibase
  10.1103/PhysRevB.87.201109} {\bibfield  {journal} {\bibinfo  {journal} {Phys.
  Rev. B}\ }\textbf {\bibinfo {volume} {87}},\ \bibinfo {pages} {201109(R)}
  (\bibinfo {year} {2013})}\BibitemShut {NoStop}%
\bibitem [{\citenamefont {Grushin}\ \emph {et~al.}(2014)\citenamefont
  {Grushin}, \citenamefont {G\'{o}mez-Le\'{o}n},\ and\ \citenamefont
  {Neupert}}]{grushin14}%
  \BibitemOpen
  \bibfield  {author} {\bibinfo {author} {\bibfnamefont {A.~G.}\ \bibnamefont
  {Grushin}}, \bibinfo {author} {\bibfnamefont {A.}~\bibnamefont
  {G\'{o}mez-Le\'{o}n}}, \ and\ \bibinfo {author} {\bibfnamefont
  {T.}~\bibnamefont {Neupert}},\ }\href {\doibase
  10.1103/PhysRevLett.112.156801} {\bibfield  {journal} {\bibinfo  {journal}
  {Phys. Rev. Lett.}\ }\textbf {\bibinfo {volume} {112}},\ \bibinfo {pages}
  {156801} (\bibinfo {year} {2014})}\BibitemShut {NoStop}%
\bibitem [{\citenamefont {{Usaj}}\ \emph {et~al.}(2014)\citenamefont {{Usaj}},
  \citenamefont {{Perez-Piskunow}}, \citenamefont {{Foa Torres}},\ and\
  \citenamefont {{Balseiro}}}]{usaj14}%
  \BibitemOpen
  \bibfield  {author} {\bibinfo {author} {\bibfnamefont {G.}~\bibnamefont
  {{Usaj}}}, \bibinfo {author} {\bibfnamefont {P.~M.}\ \bibnamefont
  {{Perez-Piskunow}}}, \bibinfo {author} {\bibfnamefont {L.~E.~F.}\
  \bibnamefont {{Foa Torres}}}, \ and\ \bibinfo {author} {\bibfnamefont
  {C.~A.}\ \bibnamefont {{Balseiro}}},\ }\href {\doibase
  10.1103/PhysRevB.90.115423} {\bibfield  {journal} {\bibinfo  {journal} {Phys.
  Rev. B}\ }\textbf {\bibinfo {volume} {90}},\ \bibinfo {pages} {115423}
  (\bibinfo {year} {2014})}\BibitemShut {NoStop}%
\bibitem [{\citenamefont {{Quelle}}\ \emph {et~al.}(2015)\citenamefont
  {{Quelle}}, \citenamefont {{Beugeling}},\ and\ \citenamefont {{Morais
  Smith}}}]{quelle15}%
  \BibitemOpen
  \bibfield  {author} {\bibinfo {author} {\bibfnamefont {A.}~\bibnamefont
  {{Quelle}}}, \bibinfo {author} {\bibfnamefont {W.}~\bibnamefont
  {{Beugeling}}}, \ and\ \bibinfo {author} {\bibfnamefont {C.}~\bibnamefont
  {{Morais Smith}}},\ }\href {\doibase 10.1016/j.ssc.2014.10.024} {\bibfield
  {journal} {\bibinfo  {journal} {Solid St. Commun.}\ }\textbf {\bibinfo
  {volume} {215}},\ \bibinfo {pages} {27} (\bibinfo {year} {2015})}\BibitemShut
  {NoStop}%
\bibitem [{\citenamefont {Thakurathi}\ \emph {et~al.}(2017)\citenamefont
  {Thakurathi}, \citenamefont {Loss},\ and\ \citenamefont
  {Klinovaja}}]{Klinovaja2017}%
  \BibitemOpen
  \bibfield  {author} {\bibinfo {author} {\bibfnamefont {M.}~\bibnamefont
  {Thakurathi}}, \bibinfo {author} {\bibfnamefont {D.}~\bibnamefont {Loss}}, \
  and\ \bibinfo {author} {\bibfnamefont {J.}~\bibnamefont {Klinovaja}},\ }\href
  {\doibase 10.1103/PhysRevB.95.155407} {\bibfield  {journal} {\bibinfo
  {journal} {Phys. Rev. B}\ }\textbf {\bibinfo {volume} {95}},\ \bibinfo
  {pages} {155407} (\bibinfo {year} {2017})}\BibitemShut {NoStop}%
\bibitem [{\citenamefont {Peralta~Gavensky}\ \emph {et~al.}(2018)\citenamefont
  {Peralta~Gavensky}, \citenamefont {Usaj},\ and\ \citenamefont
  {Balseiro}}]{Gavensky18PRB}%
  \BibitemOpen
  \bibfield  {author} {\bibinfo {author} {\bibfnamefont {L.}~\bibnamefont
  {Peralta~Gavensky}}, \bibinfo {author} {\bibfnamefont {G.}~\bibnamefont
  {Usaj}}, \ and\ \bibinfo {author} {\bibfnamefont {C.~A.}\ \bibnamefont
  {Balseiro}},\ }\href {\doibase 10.1103/PhysRevB.98.165414} {\bibfield
  {journal} {\bibinfo  {journal} {Phys. Rev. B}\ }\textbf {\bibinfo {volume}
  {98}},\ \bibinfo {pages} {165414} (\bibinfo {year} {2018})}\BibitemShut
  {NoStop}%
\bibitem [{\citenamefont {Haldane}\ and\ \citenamefont
  {Raghu}(2008)}]{Haldane:2008cc}%
  \BibitemOpen
  \bibfield  {author} {\bibinfo {author} {\bibfnamefont {F.~D.~M.}\
  \bibnamefont {Haldane}}\ and\ \bibinfo {author} {\bibfnamefont
  {S.}~\bibnamefont {Raghu}},\ }\href {\doibase 10.1103/PhysRevLett.100.013904}
  {\bibfield  {journal} {\bibinfo  {journal} {Phys. Rev. Lett.}\ }\textbf
  {\bibinfo {volume} {100}},\ \bibinfo {pages} {013904} (\bibinfo {year}
  {2008})}\BibitemShut {NoStop}%
\bibitem [{\citenamefont {Rechtsman}\ \emph {et~al.}(2013)\citenamefont
  {Rechtsman}, \citenamefont {Zeuner}, \citenamefont {Plotnik}, \citenamefont
  {Lumer}, \citenamefont {Podolsky}, \citenamefont {Dreisow}, \citenamefont
  {Nolte}, \citenamefont {Segev},\ and\ \citenamefont
  {Szameit}}]{Rechtsman:2013fe}%
  \BibitemOpen
  \bibfield  {author} {\bibinfo {author} {\bibfnamefont {M.~C.}\ \bibnamefont
  {Rechtsman}}, \bibinfo {author} {\bibfnamefont {J.~M.}\ \bibnamefont
  {Zeuner}}, \bibinfo {author} {\bibfnamefont {Y.}~\bibnamefont {Plotnik}},
  \bibinfo {author} {\bibfnamefont {Y.}~\bibnamefont {Lumer}}, \bibinfo
  {author} {\bibfnamefont {D.}~\bibnamefont {Podolsky}}, \bibinfo {author}
  {\bibfnamefont {F.}~\bibnamefont {Dreisow}}, \bibinfo {author} {\bibfnamefont
  {S.}~\bibnamefont {Nolte}}, \bibinfo {author} {\bibfnamefont
  {M.}~\bibnamefont {Segev}}, \ and\ \bibinfo {author} {\bibfnamefont
  {A.}~\bibnamefont {Szameit}},\ }\href {\doibase 10.1038/nature12066}
  {\bibfield  {journal} {\bibinfo  {journal} {Nature}\ }\textbf {\bibinfo
  {volume} {496}},\ \bibinfo {pages} {196} (\bibinfo {year}
  {2013})}\BibitemShut {NoStop}%
\bibitem [{\citenamefont {Mukherjee}\ \emph {et~al.}(2017)\citenamefont
  {Mukherjee}, \citenamefont {Spracklen}, \citenamefont {Valiente},
  \citenamefont {Andersson}, \citenamefont {{\"O}hberg}, \citenamefont
  {Goldman},\ and\ \citenamefont {Thomson}}]{Mukherjee17Ncommun}%
  \BibitemOpen
  \bibfield  {author} {\bibinfo {author} {\bibfnamefont {S.}~\bibnamefont
  {Mukherjee}}, \bibinfo {author} {\bibfnamefont {A.}~\bibnamefont
  {Spracklen}}, \bibinfo {author} {\bibfnamefont {M.}~\bibnamefont {Valiente}},
  \bibinfo {author} {\bibfnamefont {E.}~\bibnamefont {Andersson}}, \bibinfo
  {author} {\bibfnamefont {P.}~\bibnamefont {{\"O}hberg}}, \bibinfo {author}
  {\bibfnamefont {N.}~\bibnamefont {Goldman}}, \ and\ \bibinfo {author}
  {\bibfnamefont {R.~R.}\ \bibnamefont {Thomson}},\ }\href@noop {} {\bibfield
  {journal} {\bibinfo  {journal} {Nat. Commun.}\ }\textbf {\bibinfo {volume}
  {8}} (\bibinfo {year} {2017})}\BibitemShut {NoStop}%
\bibitem [{\citenamefont {J{\"o}rg}\ \emph {et~al.}(2017)\citenamefont
  {J{\"o}rg}, \citenamefont {Letscher}, \citenamefont {Fleischhauer},\ and\
  \citenamefont {von Freymann}}]{Jorg17NJP}%
  \BibitemOpen
  \bibfield  {author} {\bibinfo {author} {\bibfnamefont {C.}~\bibnamefont
  {J{\"o}rg}}, \bibinfo {author} {\bibfnamefont {F.}~\bibnamefont {Letscher}},
  \bibinfo {author} {\bibfnamefont {M.}~\bibnamefont {Fleischhauer}}, \ and\
  \bibinfo {author} {\bibfnamefont {G.}~\bibnamefont {von Freymann}},\ }\href
  {http://stacks.iop.org/1367-2630/19/i=8/a=083003} {\bibfield  {journal}
  {\bibinfo  {journal} {New Journal of Physics}\ }\textbf {\bibinfo {volume}
  {19}},\ \bibinfo {pages} {083003} (\bibinfo {year} {2017})}\BibitemShut
  {NoStop}%
\bibitem [{\citenamefont {Mukherjee}\ \emph {et~al.}(2018)\citenamefont
  {Mukherjee}, \citenamefont {Chandrasekharan}, \citenamefont {{\"O}hberg},
  \citenamefont {Goldman},\ and\ \citenamefont {Thomson}}]{Mukherjee18NCommun}%
  \BibitemOpen
  \bibfield  {author} {\bibinfo {author} {\bibfnamefont {S.}~\bibnamefont
  {Mukherjee}}, \bibinfo {author} {\bibfnamefont {H.~K.}\ \bibnamefont
  {Chandrasekharan}}, \bibinfo {author} {\bibfnamefont {P.}~\bibnamefont
  {{\"O}hberg}}, \bibinfo {author} {\bibfnamefont {N.}~\bibnamefont {Goldman}},
  \ and\ \bibinfo {author} {\bibfnamefont {R.~R.}\ \bibnamefont {Thomson}},\
  }\href {\doibase 10.1038/s41467-018-06723-y} {\bibfield  {journal} {\bibinfo
  {journal} {Nature Communications}\ }\textbf {\bibinfo {volume} {9}},\
  \bibinfo {pages} {4209} (\bibinfo {year} {2018})}\BibitemShut {NoStop}%
\bibitem [{\citenamefont {Zenesini}\ \emph {et~al.}(2009)\citenamefont
  {Zenesini}, \citenamefont {Lignier}, \citenamefont {Ciampini}, \citenamefont
  {Morsch},\ and\ \citenamefont {Arimondo}}]{zenesini09}%
  \BibitemOpen
  \bibfield  {author} {\bibinfo {author} {\bibfnamefont {A.}~\bibnamefont
  {Zenesini}}, \bibinfo {author} {\bibfnamefont {H.}~\bibnamefont {Lignier}},
  \bibinfo {author} {\bibfnamefont {D.}~\bibnamefont {Ciampini}}, \bibinfo
  {author} {\bibfnamefont {O.}~\bibnamefont {Morsch}}, \ and\ \bibinfo {author}
  {\bibfnamefont {E.}~\bibnamefont {Arimondo}},\ }\href {\doibase
  10.1103/PhysRevLett.102.100403} {\bibfield  {journal} {\bibinfo  {journal}
  {Phys. Rev. Lett.}\ }\textbf {\bibinfo {volume} {102}},\ \bibinfo {pages}
  {100403} (\bibinfo {year} {2009})}\BibitemShut {NoStop}%
\bibitem [{\citenamefont {Dalibard}\ \emph {et~al.}(2011)\citenamefont
  {Dalibard}, \citenamefont {Gerbier}, \citenamefont {Juzeli{\=u}nas},\ and\
  \citenamefont {{\"O}hberg}}]{Dalibard2011}%
  \BibitemOpen
  \bibfield  {author} {\bibinfo {author} {\bibfnamefont {J.}~\bibnamefont
  {Dalibard}}, \bibinfo {author} {\bibfnamefont {F.}~\bibnamefont {Gerbier}},
  \bibinfo {author} {\bibfnamefont {G.}~\bibnamefont {Juzeli{\=u}nas}}, \ and\
  \bibinfo {author} {\bibfnamefont {P.}~\bibnamefont {{\"O}hberg}},\ }\href
  {\doibase 10.1103/RevModPhys.83.1523} {\bibfield  {journal} {\bibinfo
  {journal} {Rev. Mod. Phys.}\ }\textbf {\bibinfo {volume} {83}},\ \bibinfo
  {pages} {1523} (\bibinfo {year} {2011})}\BibitemShut {NoStop}%
\bibitem [{\citenamefont {Aidelsburger}\ \emph {et~al.}(2011)\citenamefont
  {Aidelsburger}, \citenamefont {Atala}, \citenamefont {Nascimb{\`e}ne},
  \citenamefont {Trotzky}, \citenamefont {Chen},\ and\ \citenamefont
  {Bloch}}]{Aidelsburger:2011}%
  \BibitemOpen
  \bibfield  {author} {\bibinfo {author} {\bibfnamefont {M.}~\bibnamefont
  {Aidelsburger}}, \bibinfo {author} {\bibfnamefont {M.}~\bibnamefont {Atala}},
  \bibinfo {author} {\bibfnamefont {S.}~\bibnamefont {Nascimb{\`e}ne}},
  \bibinfo {author} {\bibfnamefont {S.}~\bibnamefont {Trotzky}}, \bibinfo
  {author} {\bibfnamefont {Y.-A.}\ \bibnamefont {Chen}}, \ and\ \bibinfo
  {author} {\bibfnamefont {I.}~\bibnamefont {Bloch}},\ }\href {\doibase
  10.1103/PhysRevLett.107.255301} {\bibfield  {journal} {\bibinfo  {journal}
  {Phys. Rev. Lett.}\ }\textbf {\bibinfo {volume} {107}},\ \bibinfo {pages}
  {255301} (\bibinfo {year} {2011})}\BibitemShut {NoStop}%
\bibitem [{\citenamefont {Struck}\ \emph {et~al.}(2011)\citenamefont {Struck},
  \citenamefont {\"Olschl\"ager}, \citenamefont {{Le Targat}}, \citenamefont
  {{Soltan-Panahi}}, \citenamefont {Eckardt}, \citenamefont {Lewenstein},
  \citenamefont {Windpassinger},\ and\ \citenamefont {Sengstock}}]{struck11}%
  \BibitemOpen
  \bibfield  {author} {\bibinfo {author} {\bibfnamefont {J.}~\bibnamefont
  {Struck}}, \bibinfo {author} {\bibfnamefont {C.}~\bibnamefont
  {\"Olschl\"ager}}, \bibinfo {author} {\bibfnamefont {R.}~\bibnamefont {{Le
  Targat}}}, \bibinfo {author} {\bibfnamefont {P.}~\bibnamefont
  {{Soltan-Panahi}}}, \bibinfo {author} {\bibfnamefont {A.}~\bibnamefont
  {Eckardt}}, \bibinfo {author} {\bibfnamefont {M.}~\bibnamefont {Lewenstein}},
  \bibinfo {author} {\bibfnamefont {P.}~\bibnamefont {Windpassinger}}, \ and\
  \bibinfo {author} {\bibfnamefont {K.}~\bibnamefont {Sengstock}},\ }\href
  {\doibase 10.1126/science.1207239} {\bibfield  {journal} {\bibinfo  {journal}
  {Science}\ }\textbf {\bibinfo {volume} {333}},\ \bibinfo {pages} {996}
  (\bibinfo {year} {2011})}\BibitemShut {NoStop}%
\bibitem [{\citenamefont {Arimondo}\ \emph {et~al.}(2012)\citenamefont
  {Arimondo}, \citenamefont {Ciampini}, \citenamefont {Eckardt}, \citenamefont
  {Holthaus},\ and\ \citenamefont {Morsch}}]{Arimondo2012}%
  \BibitemOpen
  \bibfield  {author} {\bibinfo {author} {\bibfnamefont {E.}~\bibnamefont
  {Arimondo}}, \bibinfo {author} {\bibfnamefont {D.}~\bibnamefont {Ciampini}},
  \bibinfo {author} {\bibfnamefont {A.}~\bibnamefont {Eckardt}}, \bibinfo
  {author} {\bibfnamefont {M.}~\bibnamefont {Holthaus}}, \ and\ \bibinfo
  {author} {\bibfnamefont {O.}~\bibnamefont {Morsch}},\ }\href {\doibase
  10.1016/B978-0-12-396482-3.00010-7} {\bibfield  {journal} {\bibinfo
  {journal} {Adv. At. Molec. Opt. Phys.}\ }\textbf {\bibinfo {volume} {61}},\
  \bibinfo {pages} {515} (\bibinfo {year} {2012})}\BibitemShut {NoStop}%
\bibitem [{\citenamefont {Struck}\ \emph {et~al.}(2012)\citenamefont {Struck},
  \citenamefont {\"{O}lschl\"{a}ger}, \citenamefont {Weinberg}, \citenamefont
  {Hauke}, \citenamefont {Simonet}, \citenamefont {Eckardt}, \citenamefont
  {Lewenstein}, \citenamefont {Sengstock},\ and\ \citenamefont
  {Windpassinger}}]{struck12}%
  \BibitemOpen
  \bibfield  {author} {\bibinfo {author} {\bibfnamefont {J.}~\bibnamefont
  {Struck}}, \bibinfo {author} {\bibfnamefont {C.}~\bibnamefont
  {\"{O}lschl\"{a}ger}}, \bibinfo {author} {\bibfnamefont {M.}~\bibnamefont
  {Weinberg}}, \bibinfo {author} {\bibfnamefont {P.}~\bibnamefont {Hauke}},
  \bibinfo {author} {\bibfnamefont {J.}~\bibnamefont {Simonet}}, \bibinfo
  {author} {\bibfnamefont {A.}~\bibnamefont {Eckardt}}, \bibinfo {author}
  {\bibfnamefont {M.}~\bibnamefont {Lewenstein}}, \bibinfo {author}
  {\bibfnamefont {K.}~\bibnamefont {Sengstock}}, \ and\ \bibinfo {author}
  {\bibfnamefont {P.}~\bibnamefont {Windpassinger}},\ }\href {\doibase
  10.1103/PhysRevLett.108.225304} {\bibfield  {journal} {\bibinfo  {journal}
  {Phys. Rev. Lett.}\ }\textbf {\bibinfo {volume} {108}},\ \bibinfo {pages}
  {225304} (\bibinfo {year} {2012})}\BibitemShut {NoStop}%
\bibitem [{\citenamefont {Hauke}\ \emph {et~al.}(2012)\citenamefont {Hauke},
  \citenamefont {Tieleman}, \citenamefont {Celi}, \citenamefont
  {{\"O}lschl{\"a}ger}, \citenamefont {Simonet}, \citenamefont {Struck},
  \citenamefont {Weinberg}, \citenamefont {Windpassinger}, \citenamefont
  {Sengstock}, \citenamefont {Lewenstein},\ and\ \citenamefont
  {Eckardt}}]{Hauke:2012}%
  \BibitemOpen
  \bibfield  {author} {\bibinfo {author} {\bibfnamefont {P.}~\bibnamefont
  {Hauke}}, \bibinfo {author} {\bibfnamefont {O.}~\bibnamefont {Tieleman}},
  \bibinfo {author} {\bibfnamefont {A.}~\bibnamefont {Celi}}, \bibinfo {author}
  {\bibfnamefont {C.}~\bibnamefont {{\"O}lschl{\"a}ger}}, \bibinfo {author}
  {\bibfnamefont {J.}~\bibnamefont {Simonet}}, \bibinfo {author} {\bibfnamefont
  {J.}~\bibnamefont {Struck}}, \bibinfo {author} {\bibfnamefont
  {M.}~\bibnamefont {Weinberg}}, \bibinfo {author} {\bibfnamefont
  {P.}~\bibnamefont {Windpassinger}}, \bibinfo {author} {\bibfnamefont
  {K.}~\bibnamefont {Sengstock}}, \bibinfo {author} {\bibfnamefont
  {M.}~\bibnamefont {Lewenstein}}, \ and\ \bibinfo {author} {\bibfnamefont
  {A.}~\bibnamefont {Eckardt}},\ }\href {\doibase
  10.1103/PhysRevLett.109.145301} {\bibfield  {journal} {\bibinfo  {journal}
  {Phys. Rev. Lett.}\ }\textbf {\bibinfo {volume} {109}},\ \bibinfo {pages}
  {145301} (\bibinfo {year} {2012})}\BibitemShut {NoStop}%
\bibitem [{\citenamefont {Windpassinger}\ and\ \citenamefont
  {Sengstock}(2013)}]{Windpassinger2013RPP}%
  \BibitemOpen
  \bibfield  {author} {\bibinfo {author} {\bibfnamefont {P.}~\bibnamefont
  {Windpassinger}}\ and\ \bibinfo {author} {\bibfnamefont {K.}~\bibnamefont
  {Sengstock}},\ }\href {\doibase 10.1088/0034-4885/76/8/086401} {\bibfield
  {journal} {\bibinfo  {journal} {Rep. Progr. Phys.}\ }\textbf {\bibinfo
  {volume} {76}},\ \bibinfo {pages} {086401} (\bibinfo {year}
  {2013})}\BibitemShut {NoStop}%
\bibitem [{\citenamefont {Aidelsburger}\ \emph {et~al.}(2013)\citenamefont
  {Aidelsburger}, \citenamefont {Atala}, \citenamefont {Lohse}, \citenamefont
  {Barreiro}, \citenamefont {Paredes},\ and\ \citenamefont
  {Bloch}}]{Aidelsburger:2013}%
  \BibitemOpen
  \bibfield  {author} {\bibinfo {author} {\bibfnamefont {M.}~\bibnamefont
  {Aidelsburger}}, \bibinfo {author} {\bibfnamefont {M.}~\bibnamefont {Atala}},
  \bibinfo {author} {\bibfnamefont {M.}~\bibnamefont {Lohse}}, \bibinfo
  {author} {\bibfnamefont {J.~T.}\ \bibnamefont {Barreiro}}, \bibinfo {author}
  {\bibfnamefont {B.}~\bibnamefont {Paredes}}, \ and\ \bibinfo {author}
  {\bibfnamefont {I.}~\bibnamefont {Bloch}},\ }\href {\doibase
  10.1103/PhysRevLett.111.185301} {\bibfield  {journal} {\bibinfo  {journal}
  {Phys. Rev. Lett.}\ }\textbf {\bibinfo {volume} {111}},\ \bibinfo {pages}
  {185301} (\bibinfo {year} {2013})}\BibitemShut {NoStop}%
\bibitem [{\citenamefont {Hauke}\ \emph {et~al.}(2014)\citenamefont {Hauke},
  \citenamefont {Lewenstein},\ and\ \citenamefont {Eckardt}}]{hauke13}%
  \BibitemOpen
  \bibfield  {author} {\bibinfo {author} {\bibfnamefont {P.}~\bibnamefont
  {Hauke}}, \bibinfo {author} {\bibfnamefont {M.}~\bibnamefont {Lewenstein}}, \
  and\ \bibinfo {author} {\bibfnamefont {A.}~\bibnamefont {Eckardt}},\ }\href
  {\doibase 10.1103/PhysRevLett.113.045303} {\bibfield  {journal} {\bibinfo
  {journal} {Phys. Rev. Lett.}\ }\textbf {\bibinfo {volume} {113}},\ \bibinfo
  {pages} {045303} (\bibinfo {year} {2014})}\BibitemShut {NoStop}%
\bibitem [{\citenamefont {Struck}\ \emph {et~al.}(2013)\citenamefont {Struck},
  \citenamefont {Weinberg}, \citenamefont {\"Olschl\"ager}, \citenamefont
  {Windpassinger}, \citenamefont {Simonet}, \citenamefont {Sengstock},
  \citenamefont {H\"oppner}, \citenamefont {Hauke}, \citenamefont {Eckardt},
  \citenamefont {Lewenstein},\ and\ \citenamefont {Mathey}}]{Struck:2013}%
  \BibitemOpen
  \bibfield  {author} {\bibinfo {author} {\bibfnamefont {J.}~\bibnamefont
  {Struck}}, \bibinfo {author} {\bibfnamefont {M.}~\bibnamefont {Weinberg}},
  \bibinfo {author} {\bibfnamefont {C.}~\bibnamefont {\"Olschl\"ager}},
  \bibinfo {author} {\bibfnamefont {P.}~\bibnamefont {Windpassinger}}, \bibinfo
  {author} {\bibfnamefont {J.}~\bibnamefont {Simonet}}, \bibinfo {author}
  {\bibfnamefont {K.}~\bibnamefont {Sengstock}}, \bibinfo {author}
  {\bibfnamefont {R.}~\bibnamefont {H\"oppner}}, \bibinfo {author}
  {\bibfnamefont {P.}~\bibnamefont {Hauke}}, \bibinfo {author} {\bibfnamefont
  {A.}~\bibnamefont {Eckardt}}, \bibinfo {author} {\bibfnamefont
  {M.}~\bibnamefont {Lewenstein}}, \ and\ \bibinfo {author} {\bibfnamefont
  {L.}~\bibnamefont {Mathey}},\ }\href {\doibase 10.1038/nphys2750} {\bibfield
  {journal} {\bibinfo  {journal} {Nat. Phys.}\ }\textbf {\bibinfo {volume}
  {9}},\ \bibinfo {pages} {738} (\bibinfo {year} {2013})}\BibitemShut {NoStop}%
\bibitem [{\citenamefont {Miyake}\ \emph {et~al.}(2013)\citenamefont {Miyake},
  \citenamefont {Siviloglou}, \citenamefont {Kennedy}, \citenamefont {Burton},\
  and\ \citenamefont {Ketterle}}]{Ketterle:2013}%
  \BibitemOpen
  \bibfield  {author} {\bibinfo {author} {\bibfnamefont {H.}~\bibnamefont
  {Miyake}}, \bibinfo {author} {\bibfnamefont {G.~A.}\ \bibnamefont
  {Siviloglou}}, \bibinfo {author} {\bibfnamefont {C.~J.}\ \bibnamefont
  {Kennedy}}, \bibinfo {author} {\bibfnamefont {W.~C.}\ \bibnamefont {Burton}},
  \ and\ \bibinfo {author} {\bibfnamefont {W.}~\bibnamefont {Ketterle}},\
  }\href {\doibase 10.1103/PhysRevLett.111.185302} {\bibfield  {journal}
  {\bibinfo  {journal} {Phys. Phys. Lett.}\ }\textbf {\bibinfo {volume}
  {111}},\ \bibinfo {pages} {185302} (\bibinfo {year} {2013})}\BibitemShut
  {NoStop}%
\bibitem [{\citenamefont {Anderson}\ \emph {et~al.}(2013)\citenamefont
  {Anderson}, \citenamefont {Spielman},\ and\ \citenamefont
  {Juzeli{\=u}nas}}]{Anderson2013}%
  \BibitemOpen
  \bibfield  {author} {\bibinfo {author} {\bibfnamefont {B.~M.}\ \bibnamefont
  {Anderson}}, \bibinfo {author} {\bibfnamefont {I.~B.}\ \bibnamefont
  {Spielman}}, \ and\ \bibinfo {author} {\bibfnamefont {G.}~\bibnamefont
  {Juzeli{\=u}nas}},\ }\href {\doibase 10.1103/PhysRevLett.111.125301}
  {\bibfield  {journal} {\bibinfo  {journal} {Phys. Rev. Lett.}\ }\textbf
  {\bibinfo {volume} {111}},\ \bibinfo {pages} {125301} (\bibinfo {year}
  {2013})}\BibitemShut {NoStop}%
\bibitem [{\citenamefont {Xu}\ \emph {et~al.}(2013)\citenamefont {Xu},
  \citenamefont {You},\ and\ \citenamefont {Ueda}}]{Xu2013}%
  \BibitemOpen
  \bibfield  {author} {\bibinfo {author} {\bibfnamefont {Z.-F.}\ \bibnamefont
  {Xu}}, \bibinfo {author} {\bibfnamefont {L.}~\bibnamefont {You}}, \ and\
  \bibinfo {author} {\bibfnamefont {M.}~\bibnamefont {Ueda}},\ }\href {\doibase
  10.1103/PhysRevA.87.063634} {\bibfield  {journal} {\bibinfo  {journal} {Phys.
  Rev. A}\ }\textbf {\bibinfo {volume} {87}},\ \bibinfo {pages} {063634}
  (\bibinfo {year} {2013})}\BibitemShut {NoStop}%
\bibitem [{\citenamefont {Galitski}\ and\ \citenamefont
  {Spielman}(2013)}]{Galitski2013}%
  \BibitemOpen
  \bibfield  {author} {\bibinfo {author} {\bibfnamefont {V.}~\bibnamefont
  {Galitski}}\ and\ \bibinfo {author} {\bibfnamefont {I.~B.}\ \bibnamefont
  {Spielman}},\ }\href {\doibase 10.1038/nature11841} {\bibfield  {journal}
  {\bibinfo  {journal} {Nature}\ }\textbf {\bibinfo {volume} {494}},\ \bibinfo
  {pages} {49} (\bibinfo {year} {2013})}\BibitemShut {NoStop}%
\bibitem [{\citenamefont {Goldman}\ \emph {et~al.}(2014)\citenamefont
  {Goldman}, \citenamefont {Juzeli{\=u}nas}, \citenamefont {{\"O}hberg},\ and\
  \citenamefont {Spielman}}]{Goldman2014RPP}%
  \BibitemOpen
  \bibfield  {author} {\bibinfo {author} {\bibfnamefont {N.}~\bibnamefont
  {Goldman}}, \bibinfo {author} {\bibfnamefont {G.}~\bibnamefont
  {Juzeli{\=u}nas}}, \bibinfo {author} {\bibfnamefont {P.}~\bibnamefont
  {{\"O}hberg}}, \ and\ \bibinfo {author} {\bibfnamefont {I.~B.}\ \bibnamefont
  {Spielman}},\ }\href {\doibase 10.1088/0034-4885/77/12/126401} {\bibfield
  {journal} {\bibinfo  {journal} {Rep. Progr. Phys.}\ }\textbf {\bibinfo
  {volume} {77}},\ \bibinfo {pages} {126401} (\bibinfo {year}
  {2014})}\BibitemShut {NoStop}%
\bibitem [{\citenamefont {Aidelsburger}\ \emph {et~al.}(2015)\citenamefont
  {Aidelsburger}, \citenamefont {Lohse}, \citenamefont {Schweizer},
  \citenamefont {Atala}, \citenamefont {Barreiro}, \citenamefont
  {Nascimb{\`e}ne}, \citenamefont {Cooper}, \citenamefont {Bloch},\ and\
  \citenamefont {Goldman}}]{Aidelsburger14NP}%
  \BibitemOpen
  \bibfield  {author} {\bibinfo {author} {\bibfnamefont {M.}~\bibnamefont
  {Aidelsburger}}, \bibinfo {author} {\bibfnamefont {M.}~\bibnamefont {Lohse}},
  \bibinfo {author} {\bibfnamefont {C.}~\bibnamefont {Schweizer}}, \bibinfo
  {author} {\bibfnamefont {M.}~\bibnamefont {Atala}}, \bibinfo {author}
  {\bibfnamefont {J.~T.}\ \bibnamefont {Barreiro}}, \bibinfo {author}
  {\bibfnamefont {S.}~\bibnamefont {Nascimb{\`e}ne}}, \bibinfo {author}
  {\bibfnamefont {N.~R.}\ \bibnamefont {Cooper}}, \bibinfo {author}
  {\bibfnamefont {I.}~\bibnamefont {Bloch}}, \ and\ \bibinfo {author}
  {\bibfnamefont {N.}~\bibnamefont {Goldman}},\ }\href {\doibase
  10.1038/nphys3171} {\bibfield  {journal} {\bibinfo  {journal} {Nat. Phys.}\
  }\textbf {\bibinfo {volume} {11}},\ \bibinfo {pages} {162} (\bibinfo {year}
  {2015})}\BibitemShut {NoStop}%
\bibitem [{\citenamefont {Atala}\ \emph {et~al.}(2014)\citenamefont {Atala},
  \citenamefont {Aidelsburger}, \citenamefont {Lohse}, \citenamefont
  {Barreiro}, \citenamefont {Paredes},\ and\ \citenamefont {Bloch}}]{atala14}%
  \BibitemOpen
  \bibfield  {author} {\bibinfo {author} {\bibfnamefont {M.}~\bibnamefont
  {Atala}}, \bibinfo {author} {\bibfnamefont {M.}~\bibnamefont {Aidelsburger}},
  \bibinfo {author} {\bibfnamefont {M.}~\bibnamefont {Lohse}}, \bibinfo
  {author} {\bibfnamefont {J.~T.}\ \bibnamefont {Barreiro}}, \bibinfo {author}
  {\bibfnamefont {B.}~\bibnamefont {Paredes}}, \ and\ \bibinfo {author}
  {\bibfnamefont {I.}~\bibnamefont {Bloch}},\ }\href {\doibase
  10.1038/nphys2998} {\bibfield  {journal} {\bibinfo  {journal} {Nat. Phys.}\
  }\textbf {\bibinfo {volume} {10}},\ \bibinfo {pages} {588} (\bibinfo {year}
  {2014})}\BibitemShut {NoStop}%
\bibitem [{\citenamefont {Jotzu}\ \emph {et~al.}(2014)\citenamefont {Jotzu},
  \citenamefont {Messer}, \citenamefont {Desbuquois}, \citenamefont {Lebrat},
  \citenamefont {Uehlinger}, \citenamefont {Greif},\ and\ \citenamefont
  {Esslinger}}]{Jotzu2014}%
  \BibitemOpen
  \bibfield  {author} {\bibinfo {author} {\bibfnamefont {G.}~\bibnamefont
  {Jotzu}}, \bibinfo {author} {\bibfnamefont {M.}~\bibnamefont {Messer}},
  \bibinfo {author} {\bibfnamefont {R.}~\bibnamefont {Desbuquois}}, \bibinfo
  {author} {\bibfnamefont {M.}~\bibnamefont {Lebrat}}, \bibinfo {author}
  {\bibfnamefont {T.}~\bibnamefont {Uehlinger}}, \bibinfo {author}
  {\bibfnamefont {D.}~\bibnamefont {Greif}}, \ and\ \bibinfo {author}
  {\bibfnamefont {T.}~\bibnamefont {Esslinger}},\ }\href {\doibase
  10.1038/nature13915} {\bibfield  {journal} {\bibinfo  {journal} {Nature}\
  }\textbf {\bibinfo {volume} {515}},\ \bibinfo {pages} {237} (\bibinfo {year}
  {2014})}\BibitemShut {NoStop}%
\bibitem [{\citenamefont {Zhai}(2015)}]{Zhai2014-review}%
  \BibitemOpen
  \bibfield  {author} {\bibinfo {author} {\bibfnamefont {H.}~\bibnamefont
  {Zhai}},\ }\href {\doibase 10.1088/0034-4885/78/2/026001} {\bibfield
  {journal} {\bibinfo  {journal} {Rep. Prog. Phys.}\ }\textbf {\bibinfo
  {volume} {78}},\ \bibinfo {pages} {026001} (\bibinfo {year}
  {2015})}\BibitemShut {NoStop}%
\bibitem [{\citenamefont {Kennedy}\ \emph {et~al.}(2015)\citenamefont
  {Kennedy}, \citenamefont {Burton}, \citenamefont {Chung},\ and\ \citenamefont
  {Ketterle}}]{Kennedy15}%
  \BibitemOpen
  \bibfield  {author} {\bibinfo {author} {\bibfnamefont {C.~J.}\ \bibnamefont
  {Kennedy}}, \bibinfo {author} {\bibfnamefont {W.~C.}\ \bibnamefont {Burton}},
  \bibinfo {author} {\bibfnamefont {W.~C.}\ \bibnamefont {Chung}}, \ and\
  \bibinfo {author} {\bibfnamefont {W.}~\bibnamefont {Ketterle}},\ }\href
  {\doibase 10.1038/nphys3421} {\bibfield  {journal} {\bibinfo  {journal} {Nat.
  Phys.}\ }\textbf {\bibinfo {volume} {11}},\ \bibinfo {pages} {859} (\bibinfo
  {year} {2015})}\BibitemShut {NoStop}%
\bibitem [{\citenamefont {Fl{\"a}schner}\ \emph {et~al.}(2016)\citenamefont
  {Fl{\"a}schner}, \citenamefont {Rem}, \citenamefont {Tarnowski},
  \citenamefont {Vogel}, \citenamefont {L{\"u}hmann}, \citenamefont
  {Sengstock},\ and\ \citenamefont {Weitenberg}}]{Flaschner16}%
  \BibitemOpen
  \bibfield  {author} {\bibinfo {author} {\bibfnamefont {N.}~\bibnamefont
  {Fl{\"a}schner}}, \bibinfo {author} {\bibfnamefont {B.~S.}\ \bibnamefont
  {Rem}}, \bibinfo {author} {\bibfnamefont {M.}~\bibnamefont {Tarnowski}},
  \bibinfo {author} {\bibfnamefont {D.}~\bibnamefont {Vogel}}, \bibinfo
  {author} {\bibfnamefont {D.-S.}\ \bibnamefont {L{\"u}hmann}}, \bibinfo
  {author} {\bibfnamefont {K.}~\bibnamefont {Sengstock}}, \ and\ \bibinfo
  {author} {\bibfnamefont {C.}~\bibnamefont {Weitenberg}},\ }\href {\doibase
  10.1126/science.aad4568} {\bibfield  {journal} {\bibinfo  {journal}
  {Science}\ }\textbf {\bibinfo {volume} {352}},\ \bibinfo {pages} {1091}
  (\bibinfo {year} {2016})}\BibitemShut {NoStop}%
\bibitem [{\citenamefont {Budich}\ \emph {et~al.}()\citenamefont {Budich},
  \citenamefont {Hu},\ and\ \citenamefont {Zoller}}]{Budich16}%
  \BibitemOpen
  \bibfield  {author} {\bibinfo {author} {\bibfnamefont {J.~C.}\ \bibnamefont
  {Budich}}, \bibinfo {author} {\bibfnamefont {Y.}~\bibnamefont {Hu}}, \ and\
  \bibinfo {author} {\bibfnamefont {P.}~\bibnamefont {Zoller}},\ }\href@noop {}
  {\enquote {\bibinfo {title} {Helical {Floquet} channels in {1D} lattices},}\
  }\Eprint {http://arxiv.org/abs/1608.05096} {arXiv:1608.05096} \BibitemShut
  {NoStop}%
\bibitem [{\citenamefont {Meinert}\ \emph {et~al.}(2016)\citenamefont
  {Meinert}, \citenamefont {Mark}, \citenamefont {Lauber}, \citenamefont
  {Daley},\ and\ \citenamefont {N\"agerl}}]{Nagerl2016}%
  \BibitemOpen
  \bibfield  {author} {\bibinfo {author} {\bibfnamefont {F.}~\bibnamefont
  {Meinert}}, \bibinfo {author} {\bibfnamefont {M.~J.}\ \bibnamefont {Mark}},
  \bibinfo {author} {\bibfnamefont {K.}~\bibnamefont {Lauber}}, \bibinfo
  {author} {\bibfnamefont {A.~J.}\ \bibnamefont {Daley}}, \ and\ \bibinfo
  {author} {\bibfnamefont {H.-C.}\ \bibnamefont {N\"agerl}},\ }\href {\doibase
  10.1103/PhysRevLett.116.205301} {\bibfield  {journal} {\bibinfo  {journal}
  {Phys. Rev. Lett.}\ }\textbf {\bibinfo {volume} {116}},\ \bibinfo {pages}
  {205301} (\bibinfo {year} {2016})}\BibitemShut {NoStop}%
\bibitem [{\citenamefont {{Jim{\'e}nez-Garc{\'{\i}}a}}\ \emph
  {et~al.}(2015)\citenamefont {{Jim{\'e}nez-Garc{\'{\i}}a}}, \citenamefont
  {{LeBlanc}}, \citenamefont {{Williams}}, \citenamefont {{Beeler}},
  \citenamefont {{Qu}}, \citenamefont {{Gong}}, \citenamefont {{Zhang}},\ and\
  \citenamefont {{Spielman}}}]{jimenez15}%
  \BibitemOpen
  \bibfield  {author} {\bibinfo {author} {\bibfnamefont {K.}~\bibnamefont
  {{Jim{\'e}nez-Garc{\'{\i}}a}}}, \bibinfo {author} {\bibfnamefont {L.~J.}\
  \bibnamefont {{LeBlanc}}}, \bibinfo {author} {\bibfnamefont {R.~A.}\
  \bibnamefont {{Williams}}}, \bibinfo {author} {\bibfnamefont {M.~C.}\
  \bibnamefont {{Beeler}}}, \bibinfo {author} {\bibfnamefont {C.}~\bibnamefont
  {{Qu}}}, \bibinfo {author} {\bibfnamefont {M.}~\bibnamefont {{Gong}}},
  \bibinfo {author} {\bibfnamefont {C.}~\bibnamefont {{Zhang}}}, \ and\
  \bibinfo {author} {\bibfnamefont {I.~B.}\ \bibnamefont {{Spielman}}},\ }\href
  {\doibase 10.1103/PhysRevLett.114.125301} {\bibfield  {journal} {\bibinfo
  {journal} {Phys. Rev. Lett.}\ }\textbf {\bibinfo {volume} {114}},\ \bibinfo
  {pages} {125301} (\bibinfo {year} {2015})}\BibitemShut {NoStop}%
\bibitem [{\citenamefont {{Nascimbene}}\ \emph {et~al.}(2015)\citenamefont
  {{Nascimbene}}, \citenamefont {{Goldman}}, \citenamefont {{Cooper}},\ and\
  \citenamefont {{Dalibard}}}]{nascimbene15}%
  \BibitemOpen
  \bibfield  {author} {\bibinfo {author} {\bibfnamefont {S.}~\bibnamefont
  {{Nascimbene}}}, \bibinfo {author} {\bibfnamefont {N.}~\bibnamefont
  {{Goldman}}}, \bibinfo {author} {\bibfnamefont {N.~R.}\ \bibnamefont
  {{Cooper}}}, \ and\ \bibinfo {author} {\bibfnamefont {J.}~\bibnamefont
  {{Dalibard}}},\ }\href {\doibase 10.1103/PhysRevLett.115.140401} {\bibfield
  {journal} {\bibinfo  {journal} {Phys. Rev. Lett.}\ }\textbf {\bibinfo
  {volume} {115}},\ \bibinfo {pages} {140401} (\bibinfo {year}
  {2015})}\BibitemShut {NoStop}%
\bibitem [{\citenamefont {{Perez-Piskunow}}\ \emph {et~al.}(2015)\citenamefont
  {{Perez-Piskunow}}, \citenamefont {{Foa Torres}},\ and\ \citenamefont
  {{Usaj}}}]{perez-piskunow15}%
  \BibitemOpen
  \bibfield  {author} {\bibinfo {author} {\bibfnamefont {P.~M.}\ \bibnamefont
  {{Perez-Piskunow}}}, \bibinfo {author} {\bibfnamefont {L.~E.~F.}\
  \bibnamefont {{Foa Torres}}}, \ and\ \bibinfo {author} {\bibfnamefont
  {G.}~\bibnamefont {{Usaj}}},\ }\href {\doibase 10.1103/PhysRevA.91.043625}
  {\bibfield  {journal} {\bibinfo  {journal} {Phys. Rev. A}\ }\textbf {\bibinfo
  {volume} {91}},\ \bibinfo {pages} {043625} (\bibinfo {year}
  {2015})}\BibitemShut {NoStop}%
\bibitem [{\citenamefont {{Luo}}\ \emph {et~al.}(2016)\citenamefont {{Luo}},
  \citenamefont {{Wu}}, \citenamefont {{Chen}}, \citenamefont {{Guan}},
  \citenamefont {{Gao}}, \citenamefont {{Xu}}, \citenamefont {{You}},\ and\
  \citenamefont {{Wang}}}]{Luo16Sci_Rep}%
  \BibitemOpen
  \bibfield  {author} {\bibinfo {author} {\bibfnamefont {X.}~\bibnamefont
  {{Luo}}}, \bibinfo {author} {\bibfnamefont {L.}~\bibnamefont {{Wu}}},
  \bibinfo {author} {\bibfnamefont {J.}~\bibnamefont {{Chen}}}, \bibinfo
  {author} {\bibfnamefont {Q.}~\bibnamefont {{Guan}}}, \bibinfo {author}
  {\bibfnamefont {K.}~\bibnamefont {{Gao}}}, \bibinfo {author} {\bibfnamefont
  {Z.-F.}\ \bibnamefont {{Xu}}}, \bibinfo {author} {\bibfnamefont
  {L.}~\bibnamefont {{You}}}, \ and\ \bibinfo {author} {\bibfnamefont
  {R.}~\bibnamefont {{Wang}}},\ }\href {\doibase 10.1038/srep18983} {\bibfield
  {journal} {\bibinfo  {journal} {Sci. Rep.}\ }\textbf {\bibinfo {volume}
  {6}},\ \bibinfo {pages} {18983} (\bibinfo {year} {2016})}\BibitemShut
  {NoStop}%
\bibitem [{\citenamefont {Eckardt}(2017)}]{Eckardt17RMP}%
  \BibitemOpen
  \bibfield  {author} {\bibinfo {author} {\bibfnamefont {A.}~\bibnamefont
  {Eckardt}},\ }\href@noop {} {\bibfield  {journal} {\bibinfo  {journal} {Rev.
  Mod. Phys.}\ }\textbf {\bibinfo {volume} {89}},\ \bibinfo {pages} {011004}
  (\bibinfo {year} {2017})}\BibitemShut {NoStop}%
\bibitem [{\citenamefont {Reitter}\ \emph {et~al.}(2017)\citenamefont
  {Reitter}, \citenamefont {N{\"a}ger}, \citenamefont {Wintersperger},
  \citenamefont {Str{\"a}ter}, \citenamefont {Bloch}, \citenamefont {Eckardt},\
  and\ \citenamefont {Schneider}}]{Schneider17PRL}%
  \BibitemOpen
  \bibfield  {author} {\bibinfo {author} {\bibfnamefont {M.}~\bibnamefont
  {Reitter}}, \bibinfo {author} {\bibfnamefont {J.}~\bibnamefont {N{\"a}ger}},
  \bibinfo {author} {\bibfnamefont {K.}~\bibnamefont {Wintersperger}}, \bibinfo
  {author} {\bibfnamefont {C.}~\bibnamefont {Str{\"a}ter}}, \bibinfo {author}
  {\bibfnamefont {I.}~\bibnamefont {Bloch}}, \bibinfo {author} {\bibfnamefont
  {A.}~\bibnamefont {Eckardt}}, \ and\ \bibinfo {author} {\bibfnamefont
  {U.}~\bibnamefont {Schneider}},\ }\href@noop {} {\bibfield  {journal}
  {\bibinfo  {journal} {Phys. Phys. Lett.}\ }\textbf {\bibinfo {volume}
  {119}},\ \bibinfo {pages} {200402} (\bibinfo {year} {2017})}\BibitemShut
  {NoStop}%
\bibitem [{\citenamefont {Clark}\ \emph {et~al.}(2018)\citenamefont {Clark},
  \citenamefont {Anderson}, \citenamefont {Feng}, \citenamefont {Gaj},
  \citenamefont {Levin},\ and\ \citenamefont {Chin}}]{Chin18PRL}%
  \BibitemOpen
  \bibfield  {author} {\bibinfo {author} {\bibfnamefont {L.~W.}\ \bibnamefont
  {Clark}}, \bibinfo {author} {\bibfnamefont {B.~M.}\ \bibnamefont {Anderson}},
  \bibinfo {author} {\bibfnamefont {L.}~\bibnamefont {Feng}}, \bibinfo {author}
  {\bibfnamefont {A.}~\bibnamefont {Gaj}}, \bibinfo {author} {\bibfnamefont
  {K.}~\bibnamefont {Levin}}, \ and\ \bibinfo {author} {\bibfnamefont
  {C.}~\bibnamefont {Chin}},\ }\href {\doibase 10.1103/PhysRevLett.121.030402}
  {\bibfield  {journal} {\bibinfo  {journal} {Phys. Rev. Lett.}\ }\textbf
  {\bibinfo {volume} {121}},\ \bibinfo {pages} {030402} (\bibinfo {year}
  {2018})}\BibitemShut {NoStop}%
\bibitem [{\citenamefont {Asteria}\ \emph {et~al.}()\citenamefont {Asteria},
  \citenamefont {Tran}, \citenamefont {Ozawa}, \citenamefont {Tarnowski},
  \citenamefont {Rem}, \citenamefont {Fl{\"a}schner}, \citenamefont
  {Sengstock}, \citenamefont {Goldman},\ and\ \citenamefont
  {Weitenberg}}]{Weitenberg18arXiv}%
  \BibitemOpen
  \bibfield  {author} {\bibinfo {author} {\bibfnamefont {L.}~\bibnamefont
  {Asteria}}, \bibinfo {author} {\bibfnamefont {D.~T.}\ \bibnamefont {Tran}},
  \bibinfo {author} {\bibfnamefont {T.}~\bibnamefont {Ozawa}}, \bibinfo
  {author} {\bibfnamefont {M.}~\bibnamefont {Tarnowski}}, \bibinfo {author}
  {\bibfnamefont {B.~S.}\ \bibnamefont {Rem}}, \bibinfo {author} {\bibfnamefont
  {N.}~\bibnamefont {Fl{\"a}schner}}, \bibinfo {author} {\bibfnamefont
  {K.}~\bibnamefont {Sengstock}}, \bibinfo {author} {\bibfnamefont
  {N.}~\bibnamefont {Goldman}}, \ and\ \bibinfo {author} {\bibfnamefont
  {C.}~\bibnamefont {Weitenberg}},\ }\href@noop {} {\enquote {\bibinfo {title}
  {Measuring quantized circular dichroism in ultracold topological matter},}\
  }\Eprint {http://arxiv.org/abs/arXiv:1805.11077} {arXiv:arXiv:1805.11077}
  \BibitemShut {NoStop}%
\bibitem [{\citenamefont {Shteynas}\ \emph {et~al.}()\citenamefont {Shteynas},
  \citenamefont {Lee}, \citenamefont {Top}, \citenamefont {Li}, \citenamefont
  {Jamison}, \citenamefont {Juzeli\=unas},\ and\ \citenamefont
  {Ketterle}}]{Shteynas18arXiv}%
  \BibitemOpen
  \bibfield  {author} {\bibinfo {author} {\bibfnamefont {B.}~\bibnamefont
  {Shteynas}}, \bibinfo {author} {\bibfnamefont {J.}~\bibnamefont {Lee}},
  \bibinfo {author} {\bibfnamefont {F.~C.}\ \bibnamefont {Top}}, \bibinfo
  {author} {\bibfnamefont {J.-R.}\ \bibnamefont {Li}}, \bibinfo {author}
  {\bibfnamefont {A.~O.}\ \bibnamefont {Jamison}}, \bibinfo {author}
  {\bibfnamefont {G.}~\bibnamefont {Juzeli\=unas}}, \ and\ \bibinfo {author}
  {\bibfnamefont {W.}~\bibnamefont {Ketterle}},\ }\href@noop {} {\enquote
  {\bibinfo {title} {How to dress radio-frequency photons with tunable
  momentum},}\ }\Eprint {http://arxiv.org/abs/arXiv:1807.07041}
  {arXiv:arXiv:1807.07041} \BibitemShut {NoStop}%
\bibitem [{\citenamefont {Rahav}\ \emph {et~al.}(2003)\citenamefont {Rahav},
  \citenamefont {Gilary},\ and\ \citenamefont {Fishman}}]{Rahav03}%
  \BibitemOpen
  \bibfield  {author} {\bibinfo {author} {\bibfnamefont {S.}~\bibnamefont
  {Rahav}}, \bibinfo {author} {\bibfnamefont {I.}~\bibnamefont {Gilary}}, \
  and\ \bibinfo {author} {\bibfnamefont {S.}~\bibnamefont {Fishman}},\ }\href
  {\doibase 10.1103/PhysRevA.68.013820} {\bibfield  {journal} {\bibinfo
  {journal} {Phys. Rev. A}\ }\textbf {\bibinfo {volume} {68}},\ \bibinfo
  {pages} {013820} (\bibinfo {year} {2003})}\BibitemShut {NoStop}%
\bibitem [{\citenamefont {Goldman}\ and\ \citenamefont
  {Dalibard}(2014)}]{Goldman2014}%
  \BibitemOpen
  \bibfield  {author} {\bibinfo {author} {\bibfnamefont {N.}~\bibnamefont
  {Goldman}}\ and\ \bibinfo {author} {\bibfnamefont {J.}~\bibnamefont
  {Dalibard}},\ }\href {\doibase 10.1103/PhysRevX.4.031027} {\bibfield
  {journal} {\bibinfo  {journal} {Phys. Rev. X}\ }\textbf {\bibinfo {volume}
  {4}},\ \bibinfo {pages} {031027} (\bibinfo {year} {2014})}\BibitemShut
  {NoStop}%
\bibitem [{\citenamefont {Goldman}\ \emph {et~al.}(2015)\citenamefont
  {Goldman}, \citenamefont {Dalibard}, \citenamefont {Aidelsburger},\ and\
  \citenamefont {Cooper}}]{goldman15resonant}%
  \BibitemOpen
  \bibfield  {author} {\bibinfo {author} {\bibfnamefont {N.}~\bibnamefont
  {Goldman}}, \bibinfo {author} {\bibfnamefont {J.}~\bibnamefont {Dalibard}},
  \bibinfo {author} {\bibfnamefont {M.}~\bibnamefont {Aidelsburger}}, \ and\
  \bibinfo {author} {\bibfnamefont {N.~R.}\ \bibnamefont {Cooper}},\ }\href
  {\doibase 10.1103/PhysRevA.91.033632} {\bibfield  {journal} {\bibinfo
  {journal} {Phys. Rev. A}\ }\textbf {\bibinfo {volume} {91}},\ \bibinfo
  {pages} {033632} (\bibinfo {year} {2015})}\BibitemShut {NoStop}%
\bibitem [{\citenamefont {Eckardt}\ and\ \citenamefont
  {Anisimovas}(2015)}]{Eckardt2015}%
  \BibitemOpen
  \bibfield  {author} {\bibinfo {author} {\bibfnamefont {A.}~\bibnamefont
  {Eckardt}}\ and\ \bibinfo {author} {\bibfnamefont {E.}~\bibnamefont
  {Anisimovas}},\ }\href {\doibase 10.1088/1367-2630/17/9/093039} {\bibfield
  {journal} {\bibinfo  {journal} {New J. Phys.}\ }\textbf {\bibinfo {volume}
  {17}},\ \bibinfo {pages} {093039} (\bibinfo {year} {2015})}\BibitemShut
  {NoStop}%
\bibitem [{\citenamefont {Bukov}\ \emph {et~al.}(2015)\citenamefont {Bukov},
  \citenamefont {D'Alessio},\ and\ \citenamefont {Polkovnikov}}]{Bukov2015}%
  \BibitemOpen
  \bibfield  {author} {\bibinfo {author} {\bibfnamefont {M.}~\bibnamefont
  {Bukov}}, \bibinfo {author} {\bibfnamefont {L.}~\bibnamefont {D'Alessio}}, \
  and\ \bibinfo {author} {\bibfnamefont {A.}~\bibnamefont {Polkovnikov}},\
  }\href {\doibase 10.1080/00018732.2015.1055918} {\bibfield  {journal}
  {\bibinfo  {journal} {Adv.\ Phys.}\ }\textbf {\bibinfo {volume} {64}},\
  \bibinfo {pages} {139} (\bibinfo {year} {2015})}\BibitemShut {NoStop}%
\bibitem [{\citenamefont {Itin}\ and\ \citenamefont
  {Katsnelson}(2015)}]{itin15}%
  \BibitemOpen
  \bibfield  {author} {\bibinfo {author} {\bibfnamefont {A.~P.}\ \bibnamefont
  {Itin}}\ and\ \bibinfo {author} {\bibfnamefont {M.~I.}\ \bibnamefont
  {Katsnelson}},\ }\href {\doibase 10.1103/PhysRevLett.115.075301} {\bibfield
  {journal} {\bibinfo  {journal} {Phys. Rev. Lett.}\ }\textbf {\bibinfo
  {volume} {115}},\ \bibinfo {pages} {075301} (\bibinfo {year}
  {2015})}\BibitemShut {NoStop}%
\bibitem [{\citenamefont {Mikami}\ \emph {et~al.}(2016)\citenamefont {Mikami},
  \citenamefont {Kitamura}, \citenamefont {Yasuda}, \citenamefont {Tsuji},
  \citenamefont {Oka},\ and\ \citenamefont {Aoki}}]{Mikami16PRB}%
  \BibitemOpen
  \bibfield  {author} {\bibinfo {author} {\bibfnamefont {T.}~\bibnamefont
  {Mikami}}, \bibinfo {author} {\bibfnamefont {S.}~\bibnamefont {Kitamura}},
  \bibinfo {author} {\bibfnamefont {K.}~\bibnamefont {Yasuda}}, \bibinfo
  {author} {\bibfnamefont {N.}~\bibnamefont {Tsuji}}, \bibinfo {author}
  {\bibfnamefont {T.}~\bibnamefont {Oka}}, \ and\ \bibinfo {author}
  {\bibfnamefont {H.}~\bibnamefont {Aoki}},\ }\href {\doibase
  10.1103/PhysRevB.93.144307} {\bibfield  {journal} {\bibinfo  {journal} {Phys.
  Rev. B}\ }\textbf {\bibinfo {volume} {93}},\ \bibinfo {pages} {144307}
  (\bibinfo {year} {2016})}\BibitemShut {NoStop}%
\bibitem [{\citenamefont {Heinisch}\ and\ \citenamefont
  {Holthaus}(2016)}]{Heinisch16}%
  \BibitemOpen
  \bibfield  {author} {\bibinfo {author} {\bibfnamefont {C.}~\bibnamefont
  {Heinisch}}\ and\ \bibinfo {author} {\bibfnamefont {M.}~\bibnamefont
  {Holthaus}},\ }\href {\doibase 10.1080/09500340.2016.1167263} {\bibfield
  {journal} {\bibinfo  {journal} {J. Mod. Opt.}\ }\textbf {\bibinfo {volume}
  {63}},\ \bibinfo {pages} {1768} (\bibinfo {year} {2016})}\BibitemShut
  {NoStop}%
\bibitem [{\citenamefont {Holthaus}(2016)}]{holthaus16tutorial}%
  \BibitemOpen
  \bibfield  {author} {\bibinfo {author} {\bibfnamefont {M.}~\bibnamefont
  {Holthaus}},\ }\href {\doibase 10.1088/0953-4075/49/1/013001} {\bibfield
  {journal} {\bibinfo  {journal} {J. Phys. B: At. Mol. Opt. Phys.}\ }\textbf
  {\bibinfo {volume} {49}},\ \bibinfo {pages} {013001} (\bibinfo {year}
  {2016})}\BibitemShut {NoStop}%
\bibitem [{\citenamefont {Desbuquois}\ \emph {et~al.}(2017)\citenamefont
  {Desbuquois}, \citenamefont {Messer}, \citenamefont {G{\"o}rg}, \citenamefont
  {Sandholzer}, \citenamefont {Jotzu},\ and\ \citenamefont
  {Esslinger}}]{Esslinger17PRA}%
  \BibitemOpen
  \bibfield  {author} {\bibinfo {author} {\bibfnamefont {R.}~\bibnamefont
  {Desbuquois}}, \bibinfo {author} {\bibfnamefont {M.}~\bibnamefont {Messer}},
  \bibinfo {author} {\bibfnamefont {F.}~\bibnamefont {G{\"o}rg}}, \bibinfo
  {author} {\bibfnamefont {K.}~\bibnamefont {Sandholzer}}, \bibinfo {author}
  {\bibfnamefont {G.}~\bibnamefont {Jotzu}}, \ and\ \bibinfo {author}
  {\bibfnamefont {T.}~\bibnamefont {Esslinger}},\ }\href@noop {} {\bibfield
  {journal} {\bibinfo  {journal} {Phys. Rev. A}\ }\textbf {\bibinfo {volume}
  {96}},\ \bibinfo {pages} {053602} (\bibinfo {year} {2017})}\BibitemShut
  {NoStop}%
\bibitem [{\citenamefont {Novi\v{c}enko}\ \emph {et~al.}(2017)\citenamefont
  {Novi\v{c}enko}, \citenamefont {Anisimovas},\ and\ \citenamefont
  {Juzeli\={u}nas}}]{Novicenko2017}%
  \BibitemOpen
  \bibfield  {author} {\bibinfo {author} {\bibfnamefont {V.}~\bibnamefont
  {Novi\v{c}enko}}, \bibinfo {author} {\bibfnamefont {E.}~\bibnamefont
  {Anisimovas}}, \ and\ \bibinfo {author} {\bibfnamefont {G.}~\bibnamefont
  {Juzeli\={u}nas}},\ }\href {\doibase 10.1103/PhysRevA.95.023615} {\bibfield
  {journal} {\bibinfo  {journal} {Phys. Rev. A}\ }\textbf {\bibinfo {volume}
  {95}},\ \bibinfo {pages} {023615} (\bibinfo {year} {2017})}\BibitemShut
  {NoStop}%
\bibitem [{\citenamefont {Wilczek}\ and\ \citenamefont
  {Zee}(1984)}]{Wilczek:1984}%
  \BibitemOpen
  \bibfield  {author} {\bibinfo {author} {\bibfnamefont {F.}~\bibnamefont
  {Wilczek}}\ and\ \bibinfo {author} {\bibfnamefont {A.}~\bibnamefont {Zee}},\
  }\href {\doibase 10.1103/PhysRevLett.52.2111} {\bibfield  {journal} {\bibinfo
   {journal} {Phys. Rev. Lett.}\ }\textbf {\bibinfo {volume} {52}},\ \bibinfo
  {pages} {2111} (\bibinfo {year} {1984})}\BibitemShut {NoStop}%
\bibitem [{\citenamefont {Moody}\ \emph {et~al.}(1986)\citenamefont {Moody},
  \citenamefont {Shapere},\ and\ \citenamefont {Wilczek}}]{Moody1986}%
  \BibitemOpen
  \bibfield  {author} {\bibinfo {author} {\bibfnamefont {J.}~\bibnamefont
  {Moody}}, \bibinfo {author} {\bibfnamefont {A.}~\bibnamefont {Shapere}}, \
  and\ \bibinfo {author} {\bibfnamefont {F.}~\bibnamefont {Wilczek}},\ }\href
  {\doibase 10.1103/PhysRevLett.56.893} {\bibfield  {journal} {\bibinfo
  {journal} {Phys. Rev. Lett.}\ }\textbf {\bibinfo {volume} {56}},\ \bibinfo
  {pages} {893} (\bibinfo {year} {1986})}\BibitemShut {NoStop}%
\bibitem [{\citenamefont {Zee}(1988)}]{Zee1988}%
  \BibitemOpen
  \bibfield  {author} {\bibinfo {author} {\bibfnamefont {A.}~\bibnamefont
  {Zee}},\ }\href {\doibase 10.1103/PhysRevA.38.1} {\bibfield  {journal}
  {\bibinfo  {journal} {Phys. Rev. A}\ }\textbf {\bibinfo {volume} {38}},\
  \bibinfo {pages} {1} (\bibinfo {year} {1988})}\BibitemShut {NoStop}%
\bibitem [{\citenamefont {Sambe}(1973)}]{sambe73}%
  \BibitemOpen
  \bibfield  {author} {\bibinfo {author} {\bibfnamefont {H.}~\bibnamefont
  {Sambe}},\ }\href {\doibase 10.1103/PhysRevA.7.2203} {\bibfield  {journal}
  {\bibinfo  {journal} {Phys. Rev. A}\ }\textbf {\bibinfo {volume} {7}},\
  \bibinfo {pages} {2203} (\bibinfo {year} {1973})}\BibitemShut {NoStop}%
\bibitem [{\citenamefont {Berry}(1984)}]{Berry:1984}%
  \BibitemOpen
  \bibfield  {author} {\bibinfo {author} {\bibfnamefont {M.}~\bibnamefont
  {Berry}},\ }\href {\doibase 10.1098/rspa.1984.0023} {\bibfield  {journal}
  {\bibinfo  {journal} {Proceedings of the Royal Society of London. Series A}\
  }\textbf {\bibinfo {volume} {392}},\ \bibinfo {pages} {45} (\bibinfo {year}
  {1984})}\BibitemShut {NoStop}%
\bibitem [{\citenamefont {Unanyan}\ \emph {et~al.}(1999)\citenamefont
  {Unanyan}, \citenamefont {Shore},\ and\ \citenamefont
  {Bergmann}}]{Unanyan99PRA}%
  \BibitemOpen
  \bibfield  {author} {\bibinfo {author} {\bibfnamefont {R.~G.}\ \bibnamefont
  {Unanyan}}, \bibinfo {author} {\bibfnamefont {B.~W.}\ \bibnamefont {Shore}},
  \ and\ \bibinfo {author} {\bibfnamefont {K.}~\bibnamefont {Bergmann}},\
  }\href@noop {} {\bibfield  {journal} {\bibinfo  {journal} {Phys. Rev. A}\
  }\textbf {\bibinfo {volume} {59}},\ \bibinfo {pages} {2910} (\bibinfo {year}
  {1999})}\BibitemShut {NoStop}%
\bibitem [{\citenamefont {Theuer}\ \emph {et~al.}(1999)\citenamefont {Theuer},
  \citenamefont {Unanyan}, \citenamefont {Habscheid}, \citenamefont {Klein},\
  and\ \citenamefont {Bergmann}}]{Theuer1990OE}%
  \BibitemOpen
  \bibfield  {author} {\bibinfo {author} {\bibfnamefont {H.}~\bibnamefont
  {Theuer}}, \bibinfo {author} {\bibfnamefont {R.}~\bibnamefont {Unanyan}},
  \bibinfo {author} {\bibfnamefont {C.}~\bibnamefont {Habscheid}}, \bibinfo
  {author} {\bibfnamefont {K.}~\bibnamefont {Klein}}, \ and\ \bibinfo {author}
  {\bibfnamefont {K.}~\bibnamefont {Bergmann}},\ }\href@noop {} {\bibfield
  {journal} {\bibinfo  {journal} {Opt. Express}\ }\textbf {\bibinfo {volume}
  {4}},\ \bibinfo {pages} {77} (\bibinfo {year} {1999})}\BibitemShut {NoStop}%
\bibitem [{\citenamefont {Unanyan}\ and\ \citenamefont
  {Fleischhauer}(2004)}]{Unanyan2004}%
  \BibitemOpen
  \bibfield  {author} {\bibinfo {author} {\bibfnamefont {R.~G.}\ \bibnamefont
  {Unanyan}}\ and\ \bibinfo {author} {\bibfnamefont {M.}~\bibnamefont
  {Fleischhauer}},\ }\href@noop {} {\bibfield  {journal} {\bibinfo  {journal}
  {Phys. Rev. A}\ }\textbf {\bibinfo {volume} {69}},\ \bibinfo {pages}
  {050302(R)} (\bibinfo {year} {2004})}\BibitemShut {NoStop}%
\bibitem [{\citenamefont {Ruseckas}\ \emph {et~al.}(2005)\citenamefont
  {Ruseckas}, \citenamefont {Juzeli\=unas}, \citenamefont {\"Ohberg},\ and\
  \citenamefont {Fleischhauer}}]{Ruseckas2005}%
  \BibitemOpen
  \bibfield  {author} {\bibinfo {author} {\bibfnamefont {J.}~\bibnamefont
  {Ruseckas}}, \bibinfo {author} {\bibfnamefont {G.}~\bibnamefont
  {Juzeli\=unas}}, \bibinfo {author} {\bibfnamefont {P.}~\bibnamefont
  {\"Ohberg}}, \ and\ \bibinfo {author} {\bibfnamefont {M.}~\bibnamefont
  {Fleischhauer}},\ }\href@noop {} {\bibfield  {journal} {\bibinfo  {journal}
  {Phys. Rev. Lett.}\ }\textbf {\bibinfo {volume} {95}},\ \bibinfo {pages}
  {010404} (\bibinfo {year} {2005})}\BibitemShut {NoStop}%
\bibitem [{\citenamefont {Leroux}\ \emph {et~al.}(2018)\citenamefont {Leroux},
  \citenamefont {Pandey}, \citenamefont {Rehbi}, \citenamefont {Chevy},
  \citenamefont {Miniatura}, \citenamefont {Gr{\'e}maud},\ and\ \citenamefont
  {Wilkowski}}]{Leroux2018NCommun}%
  \BibitemOpen
  \bibfield  {author} {\bibinfo {author} {\bibfnamefont {F.}~\bibnamefont
  {Leroux}}, \bibinfo {author} {\bibfnamefont {K.}~\bibnamefont {Pandey}},
  \bibinfo {author} {\bibfnamefont {R.}~\bibnamefont {Rehbi}}, \bibinfo
  {author} {\bibfnamefont {F.}~\bibnamefont {Chevy}}, \bibinfo {author}
  {\bibfnamefont {C.}~\bibnamefont {Miniatura}}, \bibinfo {author}
  {\bibfnamefont {B.}~\bibnamefont {Gr{\'e}maud}}, \ and\ \bibinfo {author}
  {\bibfnamefont {D.}~\bibnamefont {Wilkowski}},\ }\href {\doibase
  10.1038/s41467-018-05865-3} {\bibfield  {journal} {\bibinfo  {journal}
  {Nature Communications}\ }\textbf {\bibinfo {volume} {9}},\ \bibinfo {pages}
  {3580} (\bibinfo {year} {2018})}\BibitemShut {NoStop}%
\bibitem [{\citenamefont {Bohm}\ \emph {et~al.}(1992)\citenamefont {Bohm},
  \citenamefont {Kendrick}, \citenamefont {Loewe},\ and\ \citenamefont
  {Boya}}]{Bohm92JMP}%
  \BibitemOpen
  \bibfield  {author} {\bibinfo {author} {\bibfnamefont {A.}~\bibnamefont
  {Bohm}}, \bibinfo {author} {\bibfnamefont {B.}~\bibnamefont {Kendrick}},
  \bibinfo {author} {\bibfnamefont {M.}~\bibnamefont {Loewe}}, \ and\ \bibinfo
  {author} {\bibfnamefont {L.}~\bibnamefont {Boya}},\ }\href@noop {} {\bibfield
   {journal} {\bibinfo  {journal} {J. Math. Phys.}\ }\textbf {\bibinfo {volume}
  {33}},\ \bibinfo {pages} {977} (\bibinfo {year} {1992})}\BibitemShut
  {NoStop}%
\bibitem [{\citenamefont {Danilin}\ \emph {et~al.}(2018)\citenamefont
  {Danilin}, \citenamefont {Veps{\"a}l{\"a}inen},\ and\ \citenamefont
  {Paraoanu}}]{Danilin_2018}%
  \BibitemOpen
  \bibfield  {author} {\bibinfo {author} {\bibfnamefont {S.}~\bibnamefont
  {Danilin}}, \bibinfo {author} {\bibfnamefont {A.}~\bibnamefont
  {Veps{\"a}l{\"a}inen}}, \ and\ \bibinfo {author} {\bibfnamefont {G.~S.}\
  \bibnamefont {Paraoanu}},\ }\href {\doibase 10.1088/1402-4896/aab084}
  {\bibfield  {journal} {\bibinfo  {journal} {Physica Scripta}\ }\textbf
  {\bibinfo {volume} {93}},\ \bibinfo {pages} {055101} (\bibinfo {year}
  {2018})}\BibitemShut {NoStop}%
\end{thebibliography}%

\end{document}